\documentclass[twocolumn, twocolappendix]{aastex631}
\usepackage{stix, isomath, booktabs, longtable, enumitem, amsmath, multirow, threeparttable}
\let\tablenum\relax
\usepackage{siunitx}
\usepackage[version=4, arrows=pgf-filled]{mhchem}

\graphicspath{{./}{}}

\newcommand{\RJ}{\textit{R}_J}

\DeclareMathOperator{\eu}{\mathrm{e}}

\newcommand{\ra}[1]{\ang[
	angle-symbol-degree=\textsuperscript{h},
    angle-symbol-minute=\textsuperscript{m},
    angle-symbol-second=\textsuperscript{s},
    angle-symbol-over-decimal,
    ]{#1}}

\defcitealias{USGSMap}{USGS 2020}
\defcitealias{JupiterGlobe}{JPL 2001}
\defcitealias{GanymedeGlobe}{NOAA 2010}

\begin{document}

\title{Short-Timescale Spatial Variability of Ganymede's Optical Aurora}

\correspondingauthor{Zachariah Milby}
\email{zmilby@caltech.edu}

\author[0000-0001-5683-0095]{Zachariah {Milby}}
\affiliation{Division of Geological and Planetary Sciences,
California Institute of Technology}

\author[0000-0002-9068-3428]{Katherine {de~Kleer}}
\affiliation{Division of Geological and Planetary Sciences,
California Institute of Technology}

\author[0000-0002-6917-3458]{Carl {Schmidt}}
\affiliation{Center for Space Physics, Boston University}

\author[0000-0002-5548-3519]{Fran\c{c}ois {Leblanc}}
\affiliation{Laboratoire Atmosph\`{e}res, Milieux, Observations Spatiales, Centre National de la Recherche Scientifique, Sorbonne Universit\'{e}}

\begin{abstract}
Ganymede's aurora are the product of complex interactions between its intrinsic magnetosphere and the surrounding Jovian plasma environment and can be used to derive both atmospheric composition and density. In this study, we analyzed a time-series of Ganymede's optical aurora taken with Keck I/HIRES during eclipse by Jupiter on \mbox{2021-06-08 UTC}, one day after the Juno flyby of Ganymede. The data had sufficient signal-to-noise in individual 5-minute observations to allow for the first high cadence analysis of the spatial distribution of the aurora brightness and the ratio between the \qtylist{630.0;557.7}{nm} disk-integrated auroral brightnesses---a quantity diagnostic of the relative abundances of \ce{O}, \ce{O2} and \ce{H2O} in Ganymede's atmosphere. We found that the hemisphere closer to the centrifugal equator of Jupiter's magnetosphere (where electron number density is highest) was up to twice as bright as the opposing hemisphere. The dusk (trailing) hemisphere, subjected to the highest flux of charged particles from Jupiter's magnetosphere, was also consistently almost twice as bright as the dawn (leading) hemisphere. We modeled emission from simulated \ce{O2} and \ce{H2O} atmospheres during eclipse and found that if Ganymede hosts an \ce{H2O} sublimation atmosphere in sunlight, it must collapse on a faster timescale than expected to explain its absence in our data given our current understanding of Ganymede's surface properties.
\end{abstract}


\section{Introduction} \label{sec:intro}
The first detection of the presence of a tenuous atmosphere around Ganymede was made using stellar occultation measurements over half a century ago \citep{Carlson1973}. The initial occultation measurements suggested a surface pressure of \qty{0.1}{Pa}, and follow-up photo-chemical modeling showed that photolysis of \ce{H2O} and preferential escape of H would produce an \mbox{\ce{O2}-dominated} atmosphere \citep{Yung1977}. \cite{Kumar1982} showed there was an additional stable equilibrium in the \cite{Yung1977} model at a much lower surface pressure of \qty{1e-7}{Pa}, consistent with the upper limit of \qty{1e-6}{Pa} found using data taken during the \textit{Voyager 1} flyby \citep{Broadfoot1979}.

Ganymede's atmosphere is influenced by the presence of an internally-generated dipolar magnetic field offset by about \ang{10} from its rotation axis \citep{Kivelson1996}. This field redirects charged particles from Jupiter's magnetosphere towards Ganymede's planetographic poles, producing auroral ovals similar to those on Earth. The interaction between Ganymede's magnetic field and the Jovian magnetosphere in which it resides modifies its structure, in particular the planetographic latitudes on Ganymede at which the boundaries between open and closed field lines occur. Because Jupiter's magnetic field rotates with a sidereal period of around 10 hours, much faster than Ganymede's 172 hour orbital period, pressure from Jupiter's magnetosphere forces the boundary to higher latitudes on Ganymede's ram-facing trailing hemisphere \citep{Kivelson2004,Jia2009,Duling2022}.

\cite{Hall1998} reported the first detection of emission from Ganymede's atmosphere. They observed two atomic oxygen lines in the far-ultraviolet at \qtylist{130.4;135.6}{nm}, and concluded the brightness ratio of the lines meant they were produced from dissociative electron-impact on \ce{O2}. They also found the emission was spatially-confined near the satellite's poles, matching the magnetic field model of \cite{Kivelson1996}.

\cite{Feldman2000} observed the far-ultraviolet aurora on Ganymede with the Space Telescope Imaging Spectrograph (STIS) on the Hubble Space Telescope. They used a slit wider than Ganymede's angular diameter, giving them the ability to image the spatial distribution of any monochromatic emission. They detected emission at both \qtylist{130.4;135.6}{nm}, narrowly confined at latitudes above \ang{\pm 40} which was consistent with the expected location of the boundary between the open and closed field lines. They found the brightness both spatially and temporally variable, which led them to conclude the emission was auroral and driven by interactions between Ganymede's magnetosphere and plasma trapped in Jupiter's rotating magnetic field.

Ganymede's aurora are produced from the dissociation of atmospheric molecules by electrons trapped in Jupiter's magnetosphere, where rotational forces dominate the distribution of plasma. The best constraints on magnetospheric plasma properties currently available come from \textit{Galileo} and \textit{Voyager 1} data. \cite{Bagenal2011} used these data to model the space environment around Jupiter and calculate densities, energies and scale heights of electrons in the plasma sheet as a function of distance from Jupiter. \cite{Eviatar2001} analyzed the intensity of the ultraviolet observations of \cite{Feldman2000} and found that direct impact from electrons in Jupiter's rotating magnetosphere could not excite the observed aurora brightnesses given the \textit{Galileo} measurements of electron number density. They concluded that the electrons must be accelerated to higher energies by magnetospheric interactions at the open/closed field line boundary where Ganymede's magnetic field reconnects with Jupiter's magnetic field.

\cite{McGrath2013} observed differences in the morphology of the ultraviolet aurora for the leading, trailing and sub-Jovian hemispheres. These observations showed the emission at higher latitudes on the trailing/upstream hemisphere, consistent with the magneto-hydrodynamic simulations of \cite{Jia2009}. Later observations exhibit this same hemispheric morphology, demonstrating the intrinsic shape of Ganymede's auroral ovals and their spatial correlation with the open/closed field line boundary of Ganymede's magnetosphere \citep{Musacchio2017,Molyneux2018,Roth2021,Greathouse2022,Marzok2022,Saur2022}.

A major outstanding question is the composition of Ganymede's atmosphere. Most previous far-ultraviolet aurora observations \citep{Hall1998,Feldman2000,McGrath2013} concluded that the ratio of the \qtylist{130.4;135.6}{nm} aurora brightnesses were indicative of an \ce{O2} atmosphere with a column density between \qtylist{1e18;1e19}{m^{-2}}. However, \cite{Roth2021} found a difference in the ratio of the two emissions between the disk center and limb using high spatial-resolution spectra from Hubble/STIS. They attributed this variability to the presence of a localized \ce{H2O} atmosphere around the sub-solar point (near the disk center) with a peak column density of around \qty{6e19}{m^{-2}} in sunlight, which exists in addition to the global \ce{O2} atmosphere.

Though ground-based optical observations have lower spatial resolution, there are four independent optical oxygen emissions (the \SI{557.7}{nm} emission line, the \SI[parse-numbers=false]{630.0/636.4}{nm} doublet, the \SI{777.4}{nm} triplet and the \SI{844.6}{nm} triplet) compared to just two detected in the ultraviolet (the \SI{130.4}{nm} triplet and the \SI{135.6}{nm} doublet). Because optical wavelengths can be observed from the ground, large telescopes provide the ability for observing cadences with shorter integration times and better signal-to-noise. \cite{deKleer2023} published the first optical wavelength observations of Ganymede's aurora, where they found evidence for an \ce{O2} atmosphere with a column density of \qty{4.7(0.1)e18}{m^{-2}}. They found an upper-limit on the \ce{H2O} column density of \qty{3e17}{m^{-2}}, giving a maximum hemisphere-averaged $\ce{H2O}/\ce{O2}$ column density ratio of just 0.06 in eclipse. The \ce{H2O} distribution and density in the sunlit atmosphere proposed by \cite{Roth2021} would produce \qty{46}{R} of H-$\mathrm{\alpha}$ emission, whereas \cite{deKleer2023} did not detect any H-$\mathrm{\alpha}$ (in eclipse) and placed a $2\sigma$ upper limit of 1.8 R. This major difference in $\ce{H2O}$ abundance between observations is suggestive of day-night difference, which could be investigated by observing any atmospheric changes on short timescales during eclipse ingress as the satellite passes into the shadow.

To evaluate potential variability in Ganymede's atmosphere, we conducted a time-series analysis of its optical aurora using data taken on \mbox{2021-06-08 UTC}. The spatially- and temporally-averaged aurora brightnesses from this observation were published as a part of the broader data set in \cite{deKleer2023}. For this study, we examined the spatial and temporal variability between the individual observations. We analyzed changes in the hemispheric spatial distribution and brightness of the \qtylist{557.7;630.0}{nm} atomic oxygen aurora lines to evaluate evidence for any short-timescale changes in atmospheric composition. We also quantitatively evaluated whether the \ce{H2O} atmosphere modeled by \cite{Leblanc2023} to explain the \cite{Roth2021} observations would be detectable under our optical observational constraints in order to provide additional confidence in conclusions about \ce{H2O} column densities from the optical aurora. We took our optical data one day after the Juno flyby of Ganymede on \mbox{2021-06-07 UTC} which included ultraviolet observations from the onboard Ultraviolet Spectrograph (UVS) instrument \citep{Greathouse2022} and complementary HST/STIS observations just before and just after the flyby \citep{Saur2022}.

\section{Observations and Data Reduction}

\begin{figure}
\includegraphics[width=\columnwidth]{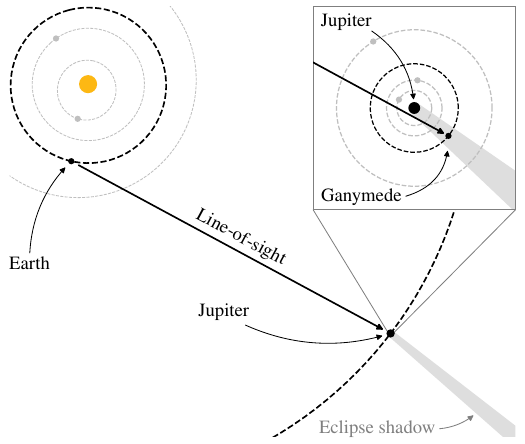}
\caption{Line-of-sight observing geometry of Ganymede during eclipse by Jupiter on \mbox{2021-06-08 UTC}. The inset axis shows a zoom-in on the Jovian system with the positions and orbits of the other Galilean satellites. Earth's position relative to Jupiter allowed us to observe Ganymede as it passed into Jupiter's umbra. We retrieved the positions of planets and their orbital elements using JPL Horizons, and we have exaggerated the physical sizes of the planets and Jupiter's umbra (but not the relative spacing of the orbits) for illustrative purposes.}
\label{fig:solar-system-geometry}
\end{figure}

We analyzed 17 spectra of Ganymede in eclipse (see table \ref{tab:observation-information}), taken on \mbox{2021-06-08 UTC} using the High Resolution Echelle Spectrometer \citep[HIRES,][]{Vogt1994} on the Keck I telescope at the summit of Maunakea. Average seeing over the course of the eclipse observations was about \ang{;;0.55}.\footnote{Historical seeing conditions for Maunakea are available at \url{http://mkwc.ifa.hawaii.edu/current/seeing/}.} Ganymede's average velocity relative to Earth was about \qty{-24}{km.s^{-1}} (the negative sign indicating motion toward Earth), a velocity sufficient to Doppler-shift Ganymede's monochromatic auroral emission from telluric emission line counterparts. Between 12:48 and \mbox{16:15 UTC}, Ganymede passed through Jupiter's umbra, allowing observation of the faint auroral emissions from its atmosphere without the overwhelming presence of reflected solar continuum. Though the full set of eclipse observations includes 17 spectra, we analyzed only 15; we eliminated the last two spectra (taken during nautical twilight) due to large systematic contamination of scattered light from Earth's atmosphere as the Sun rose. The partial umbral eclipse lasted for about 8 minutes at the beginning and end of the umbral eclipse. The positions of Earth and Jupiter relative to the Sun allowed the telescope's line-of-sight to see Ganymede enter Jupiter's umbra just beyond Jupiter's eastern limb (see figure \ref{fig:solar-system-geometry} for a graphical depiction of this viewing geometry). We used the JPL Horizons Ephemeris Service\footnote{Web interface available at \url{https://ssd.jpl.nasa.gov/horizons/app.html}. We used the Python interface provided through Astroquery (\url{https://astroquery.readthedocs.io/en/latest/jplhorizons/jplhorizons.html}).} (hereafter called JPL Horizons) to determine when Ganymede would be eclipsed by Jupiter and observable from Maunakea.

\begin{figure}
\includegraphics[width=\columnwidth]{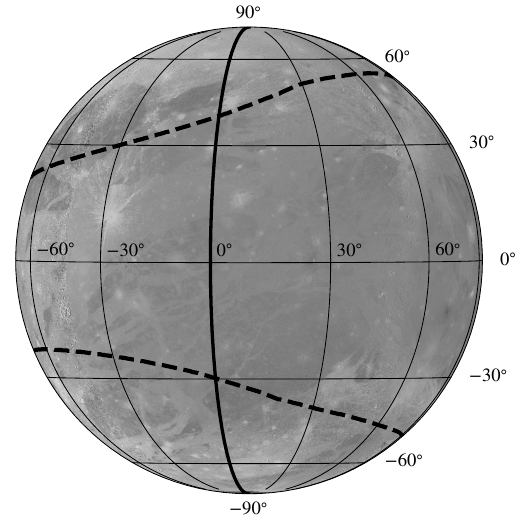}
\caption{Hemispheric observing geometry of Ganymede on 2021-06-08 between 12:58 and \mbox{15:08 UTC}. Due to the geometry of eclipse observations, the line-of-sight intersected near the center of the sub-Jovian hemisphere. The sub-observer longitude changed over the duration of the observations by \ang[angle-symbol-over-decimal=false]{\pm2.3} from the central longitude in this projection. Labels mark latitude and planetocentric east longitude; the thicker black line shows the prime meridian. The dashed black lines show the locations of the auroral ovals \citep{Duling2022}, marking the boundary between the open and closed field lines of Ganymede's magnetic field. In this view the leading hemisphere is on the left and the trailing hemisphere is on the right. Ganymede surface map courtesy of the United States Geological Survey \citepalias{USGSMap}.}
\label{fig:observing-geometry}
\end{figure}

Observations of Ganymede as it passes through Jupiter's shadow exclusively measure the sub-Jovian hemisphere. The observations in this data set had a sub-observer east longitude between \ang[angle-symbol-over-decimal=false]{8.7} and \ang[angle-symbol-over-decimal=false]{13.3}. Figure \ref{fig:observing-geometry} shows an average view of Ganymede as observed on \mbox{2021-06-08 UTC}. (The sub-observer longitude only varied from this projection by \ang[angle-symbol-over-decimal=false]{\pm2.3} over the duration of the observations.)

Twilight on the summit of Maunakea began at \mbox{14:18 UTC} and astronomical twilight ended at \mbox{14:49 UTC}. Consequently, the last two observations exhibit greater uncertainty due to higher background from scattered sunlight. Sunrise occurred at \mbox{15:46 UTC}, preventing any observations of the end of the eclipse.

The HIRES instrument allows an observer to change the angles of the echelle and cross-disperser gratings in order to optimize wavelength coverage on the detectors. The setup chosen for these observations includes emission for atomic oxygen at \qtylist{557.7;630.0;636.4;777.4;844.6}{nm} along with H-$\mathrm{\alpha}$ at \qty{656.3}{nm}. \cite{deKleer2023} reported the first analysis of these data, evaluating disk-integrated auroral brightnesses retrieved from spectra averaged over all individual observations. The brighter oxygen aurora emissions at \qtylist{557.7;630.0}{nm} have sufficient signal-to-noise in the individual five-minute exposures to allow for an analysis of both spatial and temporal variability, which we will present in this study.

In addition to the eclipse spectra of Ganymede, our data included 10 bias exposures, 4 flat lamp exposures and 5 thorium-argon (\ce{ThAr}) arc lamp exposures taken a few hours prior to the eclipse. For flux calibration, we took a spectrum of Jupiter's central meridian with the slit oriented north-south in the center of the disk. Because Ganymede was eclipsed by Jupiter and not visible for guiding, we used JPL Horizons to calculate offsets from another nearby Galilean satellite (the ``guide'' satellite) and tracking rates in right ascension and declination. We then slewed the telescope manually from the guide satellite to the expected position of Ganymede and took five-minute exposures with manual tracking rates. After each exposure, we offset back to the expected position of another nearby Galilean satellite and took a fiducial exposure to record the expected position of Ganymede within the slit. The guide satellite was initially Io, but after two exposures we switched to Europa because Io began transiting Jupiter's disk.

\subsection{HIRES Data Reduction Pipeline}
We reduced the data using an improved version of the pipeline described in \cite{deKleer2023}. The pipeline exists in two parts: the data reduction pipeline \citep{drpzenodo} is a generic HIRES pipeline, while the data calibration/brightness retrieval pipeline \citep{aurorazenodo} is specific to the aurora observations. For each calculation the data reduction and flux calibration pipelines propagate errors as described below. We used version 2.1.0 of the data reduction pipeline and version 2.15.0 of the data calibration/brightness retrieval pipeline.

\subsubsection{Reduction}
We designed our data reduction pipeline to work with data taken both before the 2004 detector upgrade (the single $2048\times2048$ pixel detector, hereafter called ``legacy'' data) and after the detector upgrade (the current three-detector mosaic setup, hereafter called ``mosaic'' data). The data reduction pipeline first combines the mosaic images into a single image with proper physical separation so that orders crossing between the detectors can still be partially used (this step is skipped for the legacy data). It uses a standard star observation (or similar bright point source) to find traces along the spectral dimension for each echelle order. Using the pixel positions of the traces, it constructs an order mask image using the slit length and pixel dimensions, both in units of arcseconds. This mask is an array of zeros everywhere except half of the length of the slit above and below the traces, which are set to ones. To find the edges of the orders, it cross-correlates the mask image with the master flat-field along the spatial axis, which is able to account for the effect of overlapping orders, orders crossing between detectors and a reference trace which isn't precisely centered along the spatial axis. It extracts and rectifies individual orders by taking each pixel along the spatial dimension which fell within the boundaries of the mask order at the maximum offset of the cross correlation.

To produce a wavelength solution, the pipeline takes the rectified \ce{ThAr} arc lamp spectra, averages them along the spatial axis (to produce a one-dimensional spectrum) and normalizes them. It then produces a two-dimensional ``image'' of each one-dimensional spectrum stacked together. The HIRES data reduction pipeline MAKEE\footnote{\url{https://sites.astro.caltech.edu/~tb/makee/}} includes templates taken at a variety of echelle and cross-disperser angles which identify order numbers, wavelengths and central pixel positions of lines in the thorium-argon arc lamp spectra. Our pipeline finds the template closest to the echelle and cross-disperser angles used in the observations, then constructs a similar two-dimensional template of stacked one-dimensional spectra. It simulates individual lines within the one-dimensional spectra by using a Gaussian line profile at the central pixel position with a full-width at half-maximum (FWHM) equal to the slit width. It then cross-correlates the template reference spectrum with the observed arc lamp spectrum along both the spatial and spectral axes. It uses the maximum cross-correlation to construct an initial wavelength solution. It then fits a Gaussian function to the observed spectrum at the initial pixel position and assigns the center of the best-fit as the fractional pixel position corresponding to the wavelength guess. After assigning refined pixel positions to every identified wavelength, the pipeline fits a two-dimensional polynomial surface to construct a complete wavelength solution for each spectral-dimension pixel in each order. The use of lines identified in adjacent orders allows for a better wavelength solution in orders with fewer identified lines. Finally, it reduces the data by subtracting a median bias and flat-fielding using a normalized median flat-field, then corrects for wavelength-dependent airmass-extinction based on the median curve from \cite{Buton2013}, removing the diluting effects of Earth's atmosphere from all science images, including flux calibration images of Jupiter's central meridian (described in the following section).

\subsubsection{Flux Calibration}
The calibration pipeline uses a spectrum of Jupiter's central meridian taken on the same night as the eclipse observations, a solar radiance reference spectrum $F_\mathrm{E}$ at \qty{1}{au} \citep{Coddington2023} and Jupiter's spectral reflectivity $(I/F)$ from  \cite{Woodman1979} to calibrate the data from [\unit{counts.s^{-1}}] to rayleighs [R], a unit of photon column emission commonly used for aurora and airglow, defined as \mbox{$\qty{1}{R} \equiv \qty[parse-numbers=false]{(10^{10}/4\mathrm{\pi})}{ph.s^{-1}.m^{-2}.sr^{-1}}$}.

For Ganymede (and the other icy satellites Europa and Callisto), the pipeline retrieves aurora brightnesses for atomic oxygen emission at \qtylist{557.7;630.0;636.4;777.4;844.6}{nm} and for atomic hydrogen emission at \qty{656.3}{nm}. To calculate the expected spectral brightness $B_\mathrm{J}$ of Jupiter, it scales the solar reference spectrum with units of $[\unit{R.nm^{-1}}]$ by the square of the distance between the Sun and Jupiter $a_\mathrm{J}$ at the time of the observation and applies Jupiter's wavelength-dependent spectral reflectivity:
\begin{equation}
    B_\mathrm{J} = F_\mathrm{E}\left(\frac{1\,\mathrm{au}}{a_\mathrm{J}}\right)^2 \left(\frac{I}{F}\right).
\end{equation}

Jupiter fills the slit, so to determine the observed flux rate $\dot{N}_\mathrm{J}$ from Jupiter at a particular wavelength $\lambda$, the pipeline calculates the median value for the column in the two-dimensional Jupiter meridian spectrum containing the desired wavelength, then multiplies that median value by the number of pixels subtended by the slit on the detector, thereby estimating the total count rate at a given wavelength $\lambda$ from the slit.

Individual HIRES mosaic detector pixels have angular dimensions of $\ang{;;0.119}\times\ang{;;0.179}$ along the spatial and spectral dimensions. The data in this study have $3\times 1$ spatial/spectral binning, so each bin has a projected size of $\ang{;;0.358}\times\ang{;;0.179}$ on the sky, or an angular area of about \qty{0.0641}{arcsec^2} (\qty{1.51e-12}{sr}). HIRES users can choose from a series of slit length and width combinations by means of a series of deckers (movable metal plates containing the slits). The D3 decker used for these observations has a projected angular size of $\ang{;;1.722}\times\ang{;;7}$, so the entrance area corresponds to 188 bins on the detector.

For each individual observation of Ganymede, the pipeline takes the section of the two-dimensional spectrum within \qty{\pm 0.25}{nm} for targeted auroral wavelength (Doppler-shifted by Ganymede's velocity relative to Earth), then produces calibrated two-dimensional images by multiplying the ratio between the observed count rate per bin from Ganymede to the observed count rate from Jupiter by the physical unit conversion factor $B_\mathrm{J}$ described above,
\begin{equation}
    B_\mathrm{G} = B_\mathrm{J} \left(\frac{\dot{N}_\mathrm{G}}{\mathrm{\Omega}_\mathrm{bin}}\right) \left(\frac{\mathrm{\Omega_\mathrm{slit}}}{\dot{N}_\mathrm{J}}\right)w_\mathrm{slit}\mathrm{\Delta}\lambda,
\end{equation}
where $\dot{N}_\mathrm{G}$ is the observed flux rate from Ganymede in [\unit{counts.s^{-1}.bin^{-1}}], $\mathrm{\Omega}_\mathrm{bin}$ is the solid angular size of one detector bin in [\unit{sr.bin^{-1}}], $\dot{N}_\mathrm{J}$ is the observed flux rate from Jupiter in [\unit{counts.s^{-1}}] calculated as described above and $\mathrm{\Omega}_\mathrm{slit}$ is the solid angular size of the slit in [sr]. To account for the spectral resolution of the slit, it multiplies by the width of the slit $w_\mathrm{slit}$ in [bins] and the wavelength dispersion $\mathrm{\Delta}\lambda$ in [\unit{nm.bin^{-1}}] at the targeted wavelength.

To calculate disk-integrated brightnesses for the science target satellite, it averages the emission over a user-defined circular aperture $\mathrm{\Omega}_a$ which is larger than the apparent angular size of the target. Because it assumes emission from a disk with the angular area of Ganymede $\mathrm{\Omega}_\mathrm{G}$, it scales the average by the ratio $(\mathrm{\Omega}_a/\mathrm{\Omega}_\mathrm{G})^2$.

\section{Analysis}

Our analysis makes use of the right-handed equivalent of the System III coordinate system (with longitudes measured positively to the east rather than to the west). In this frame, the Joviographic rotation axis $\mathbf{\Omega}$ defines latitude $\lambda_\textsc{iii}$ and the rotation of Jupiter's magnetic field defines longitude $\phi_\textsc{iii,rh}$ such that the magnetic field's rotation axis is offset from the Joviographic rotation axis by \ang[angle-symbol-over-decimal=false]{9.5} toward \ang{159} longitude \citep{Connerney1998}. We also calculated magnetic latitudes $\lambda_\mathrm{m}$ defined by the magnetic rotation axis $\mathbf{\Omega}_\mathrm{m}$ by converting System III Joviographic latitude $\lambda_\textsc{iii}$ to the magnetospheric reference frame using \mbox{$\lambda_\mathrm{m} = \ang[angle-symbol-over-decimal=false]{9.5} \cos(\phi_\textsc{iii,rh} - \ang{159}) - \lambda_\textsc{iii}$}. Longitudes in the magnetospheric coordinate system are the same as the right-handed System III Joviographic coordinate system \mbox{($\phi_\mathrm{m} = \phi_\textsc{iii,rh}$)}. 

\subsection{Retrieval of Disk-Integrated Brightnesses}\label{sec:systematics}

Table \ref{tab:observation-information} lists the observation parameters and retrieved brightnesses for each of the 17 eclipse spectra in the time series. The apparent emission covers an area on the detector larger than the size of Ganymede's apparent disk because of the blurring effect of atmospheric seeing and telescope pointing variability between observations. To determine an appropriate aperture size, we used the two-dimensional spectra containing the bright \qty{630.0}{nm} emission. We found an aperture with a radius of \ang{;;1.75} fully enclosed the apparent flux, so we used this aperture size for all of the different spectral lines. This was about $2.25\times$ the size of Ganymede's apparent angular radius.

\begin{table*}
\centering
\caption{Overview of the Keck/HIRES observations of Ganymede in eclipse on \mbox{2021-06-08 UTC}.}
\label{tab:observation-information}
\begin{tabular}{cccccS[table-format=1.3,table-number-alignment=right]S[table-format=2.1,table-number-alignment=right]S[table-format=+1.3,table-number-alignment=right]S[table-format=3.1,table-number-alignment=right]S[table-format=2.3,table-number-alignment=right]S[table-format=2.1,table-number-alignment=right]S[table-format=+1.3,table-number-alignment=right]S[table-format=+2.1,table-number-alignment=right]}
\toprule
 {$t_\mathrm{E}$\textsuperscript{\scriptsize{a}}} \relax& {$t_\mathrm{LT}$\textsuperscript{\scriptsize{b}}} \relax&  {Right Ascension} \relax&
 {Declination} \relax& \multirow{2}{*}{Airmass} \relax& {$\lambda_\mathrm{obs}$\textsuperscript{\scriptsize{c}}} \relax& {$\phi_\mathrm{obs}$\textsuperscript{\scriptsize{d}}} \relax& {$\lambda_\mathrm{m}$\textsuperscript{\scriptsize{e}}} \relax& {$\phi_\mathrm{m}$\textsuperscript{\scriptsize{f}}} \relax& {$r_\mathrm{G}$\textsuperscript{\scriptsize{g}}} \relax& {$\theta$\textsuperscript{\scriptsize{h}}} \relax& {$d$\textsuperscript{\scriptsize{i}}} \relax& {$v_\mathrm{rel}$\textsuperscript{\scriptsize{j}}}\\
 {[UTC]} \relax& {[UTC]} \relax& {[HMS]} \relax& [DMS] \relax& \relax& {[deg]} \relax& {[deg]} \relax& {[deg]} \relax& {[deg]} \relax& {[\unit{\RJ}]}  \relax& {[arcsec]} \relax& {[\unit{\RJ}]} \relax& {[\unit{km.s^{-1}}]}\\
\midrule
12:58:16 \relax& 12:19:22 \relax& \ra{22;15;54.230} \relax& \ang{-11;44;25.152} \relax& 1.519 \relax& 0.564 \relax& 11.9 \relax& 7.014 \relax& 115.3 \relax& 14.936 \relax& 78.4 \relax& 1.555 \relax& -23.7\\
13:05:29 \relax& 12:26:35 \relax& \ra{22;15;54.358} \relax& \ang{-11;44;24.432} \relax& 1.479 \relax& $\vdots$ \relax& 11.7 \relax& 6.525 \relax& 111.2 \relax& $\vdots$ \relax& 77.0 \relax& 1.446 \relax& -23.8\\
13:22:12 \relax& 12:43:18 \relax& \ra{22;15;54.650} \relax& \ang{-11;44;22.812} \relax& 1.401 \relax& $\vdots$ \relax& 11.1 \relax& 5.273 \relax& 101.7 \relax& $\vdots$ \relax& 73.9 \relax& 1.164 \relax& -23.8\\
13:29:41 \relax& 12:50:47 \relax& \ra{22;15;54.782} \relax& \ang{-11;44;22.092} \relax& 1.371 \relax& $\vdots$ \relax& 10.8 \relax& 4.664 \relax& 97.4 \relax& $\vdots$ \relax& 72.5 \relax& 1.027 \relax& -23.9\\
13:36:46 \relax& 12:57:52 \relax& \ra{22;15;54.907} \relax& \ang{-11;44;21.408} \relax& 1.346 \relax& $\vdots$ \relax& 10.6 \relax& 4.065 \relax& 93.4 \relax& $\vdots$ \relax& 71.2 \relax& 0.892 \relax& -23.9\\
13:44:33 \relax& 13:05:39 \relax& \ra{22;15;55.046} \relax& \ang{-11;44;20.652} \relax& 1.320 \relax& $\vdots$ \relax& 10.3 \relax& 3.385 \relax& 89.0 \relax& $\vdots$ \relax& 69.7 \relax& 0.738 \relax& -24.0\\
13:51:41 \relax& 13:12:47 \relax& \ra{22;15;55.171} \relax& \ang{-11;44;19.932} \relax& 1.299 \relax& $\vdots$ \relax& 10.0 \relax& 2.744 \relax& 84.9 \relax& $\vdots$ \relax& 68.4 \relax& 0.592 \relax& -24.0\\
13:58:52 \relax& 13:19:58 \relax& \ra{22;15;55.298} \relax& \ang{-11;44;19.248} \relax& 1.280 \relax& $\vdots$ \relax& 9.8 \relax& 2.085 \relax& 80.8 \relax& $\vdots$ \relax& 67.0 \relax& 0.443 \relax& -24.0\\
14:08:09 \relax& 13:29:15 \relax& \ra{22;15;55.462} \relax& \ang{-11;44;18.348} \relax& 1.257 \relax& $\vdots$ \relax& 9.5 \relax& 1.219 \relax& 75.5 \relax& $\vdots$ \relax& 65.3 \relax& 0.246 \relax& -24.1\\
14:15:28 \relax& 13:36:34 \relax& \ra{22;15;55.591} \relax& \ang{-11;44;17.628} \relax& 1.242 \relax& $\vdots$ \relax& 9.2 \relax& 0.530 \relax& 71.4 \relax& $\vdots$ \relax& 63.9 \relax& 0.089 \relax& -24.1\\
14:22:45 \relax& 13:43:51 \relax& \ra{22;15;55.718} \relax& \ang{-11;44;16.908} \relax& 1.228 \relax& $\vdots$ \relax& 9.0 \relax& -0.158 \relax& 67.2 \relax& $\vdots$ \relax& 62.5 \relax& -0.068 \relax& -24.1\\
14:29:56 \relax& 13:51:02 \relax& \ra{22;15;55.846} \relax& \ang{-11;44;16.188} \relax& 1.216 \relax& $\vdots$ \relax& 8.7 \relax& -0.835 \relax& 63.1 \relax& $\vdots$ \relax& 61.2 \relax& -0.222 \relax& -24.2\\
14:37:35 \relax& 13:58:41 \relax& \ra{22;15;55.980} \relax& \ang{-11;44;15.468} \relax& 1.205 \relax& $\vdots$ \relax& 8.4 \relax& -1.550 \relax& 58.8 \relax& $\vdots$ \relax& 59.7 \relax& -0.384 \relax& -24.2\\
14:44:47 \relax& 14:05:53 \relax& \ra{22;15;56.107} \relax& \ang{-11;44;14.748} \relax& 1.197 \relax& $\vdots$ \relax& 8.2 \relax& -2.214 \relax& 54.7 \relax& $\vdots$ \relax& 58.4 \relax& -0.535 \relax& -24.2\\
14:52:33 \relax& 14:13:39 \relax& \ra{22;15;56.246} \relax& \ang{-11;44;13.992} \relax& 1.189 \relax& $\vdots$ \relax& 7.9 \relax& -2.917 \relax& 50.2 \relax& $\vdots$ \relax& 56.9 \relax& -0.695 \relax& -24.3\\
15:01:18 \relax& 14:22:24 \relax& \ra{22;15;56.400} \relax& \ang{-11;44;13.128} \relax& 1.181 \relax& $\vdots$ \relax& 7.6 \relax& -3.688 \relax& 45.3 \relax& $\vdots$ \relax& 55.2 \relax& -0.869 \relax& -24.3\\
15:08:29 \relax& 14:29:35 \relax& \ra{22;15;56.527} \relax& \ang{-11;44;12.408} \relax& 1.177 \relax& $\vdots$ \relax& 7.4 \relax& -4.299 \relax& 41.2 \relax& $\vdots$ \relax& 53.9 \relax& -1.007 \relax& -24.3\\
\midrule
\multicolumn{2}{l}{Average} \relax& \ra{22;15;55.355} \relax& \ang{-11;44;18.780} \relax& 1.289 \relax& 0.564 \relax& 9.8 \relax& 1.989 \relax& 81.0 \relax& 14.936 \relax& 65.6 \relax& 0.419 \relax& -24.0\\
\bottomrule
\multicolumn{13}{p{0.935\linewidth}}{\textbf{Notes:} The average values and analyses in this paper do not include the last two observations at 15:01:18 and 15:08:29 which were taken during nautical twilight and exhibited significant scattered light contribution from Earth's atmosphere, affecting retrieved brightnesses and background subtraction.}\\[-5pt]
\multicolumn{13}{p{0.92\linewidth}}{
\begin{itemize}[nosep, labelsep=-1.5pt, align=left, leftmargin=*]
\item[\textsuperscript{a}]UTC time on Earth at the start of the observation.
\item[\textsuperscript{b}]UTC time on Earth corrected for light-travel time between Ganymede and Maunakea.
\item[\textsuperscript{c}]Sub-observer latitude as observed from Maunakea.
\item[\textsuperscript{d}]Sub-observer east longitude on Ganymede as observed from Maunakea.
\item[\textsuperscript{e}]Magnetospheric latitude of Ganymede.
\item[\textsuperscript{f}]Magnetospheric longitude of Ganymede (the same as the System III west longitude, converted here to east longitude).
\item[\textsuperscript{g}]Ganymede's orbital distance from Jupiter.
\item[\textsuperscript{h}]Disk-center-to-disk-center angular separation between the Jupiter and Ganymede. Jupiter's angular radius was \ang{;;21.11} on \mbox{2021-06-08 UTC}, so the effective separation from Jupiter's limb is smaller than the listed value.
\item[\textsuperscript{i}]Distance between Ganymede and the plasma sheet centrifugal equator; positive when Ganymede is above the mid-plane and negative when Ganymede is below the mid-plane.
\item[\textsuperscript{j}]Velocity of Ganymede relative to an observer on Earth (the negative sign indicates motion toward the observer).
\end{itemize}
}
\end{tabular}
\end{table*}

\begin{table*}
\centering
\caption{Retrieved optical aurora brightnesses for emission from atomic oxygen and atomic hydrogen.}
\label{tab:disk-integrated-brightness}
\begin{tabular}{lS[table-format=2.1(2),table-number-alignment=right, separate-uncertainty]S[table-format=3.1(1.1),table-number-alignment=right, separate-uncertainty]S[table-format=2.1(1.1),table-number-alignment=right, separate-uncertainty]S[table-format=+2(2),table-number-alignment=right, separate-uncertainty]S[table-format=+2.1(1.1),table-number-alignment=right, separate-uncertainty, table-align-uncertainty]S[table-format=+2.1(2),table-number-alignment=right, separate-uncertainty, table-align-uncertainty]}
\toprule
& \multicolumn{6}{c}{{Disk-Integrated Brightness}}\\
\cmidrule(lr){2-7}
\multirow[t]{2}{*}{\shortstack[l]{Observation\\Start Time}} \relax& {\qty{557.7}{nm} [O\,\textsc{i}]} \relax& {\qty{630.0}{nm} [O\,\textsc{i}]} \relax& {\qty{636.4}{nm} [O\,\textsc{i}]} \relax& {\qty{656.3}{nm} H\,\textsc{i}} \relax& {\qty{777.4}{nm} O\,\textsc{i}} \relax& {\qty{844.6}{nm} O\,\textsc{i}}\\
 \multicolumn{1}{c}{[UTC]} \relax& {[R]} \relax& {[R]} \relax& {[R]} \relax& {[R]} \relax& {[R]} \relax& {[R]}\\
\midrule
12:58:16 \relax& 13(2) \relax& 125(12) \relax& 51(5) \relax& 9(9) \relax& 47(6) \relax& 14(5)\\
13:05:29 \relax& 16(2) \relax& 102(10) \relax& 42(4) \relax& -15(9) \relax& 15(5) \relax& -4(4)\\
13:22:12 \relax& 12(2) \relax& 124(12) \relax& 50(5) \relax& 34(9) \relax& 25(5) \relax& 13(4)\\
13:29:41 \relax& 8(2) \relax& 132(12) \relax& 49(5) \relax& -2(9) \relax& 32(5) \relax& 25(5)\\
13:36:46 \relax& 11(2) \relax& 121(11) \relax& 44(5) \relax& -23(9) \relax& 33(5) \relax& 14(4)\\
13:44:33 \relax& 8(2) \relax& 151(14) \relax& 48(5) \relax& -2(9) \relax& 47(6) \relax& 15(4)\\
13:51:41 \relax& 14(2) \relax& 143(13) \relax& 47(5) \relax& 9(9) \relax& 13(4) \relax& 36(5)\\
13:58:52 \relax& 10(2) \relax& 113(11) \relax& 35(4) \relax& 11(9) \relax& 24(5) \relax& 11(4)\\
14:08:09 \relax& 12(2) \relax& 122(11) \relax& 43(4) \relax& -5(9) \relax& 14(5) \relax& 17(5)\\
14:15:28 \relax& 2(2) \relax& 118(11) \relax& 33(4) \relax& 0(10) \relax& 6(4) \relax& 6(7)\\
14:22:45 \relax& 19(3) \relax& 131(12) \relax& 41(4) \relax& 21(10) \relax& 31(5) \relax& 36(6)\\
14:29:56 \relax& 16(3) \relax& 150(14) \relax& 43(4) \relax& -14(9) \relax& 20(5) \relax& 23(6)\\
14:37:35 \relax& 18(3) \relax& 156(14) \relax& 47(5) \relax& 7(10) \relax& 35(6) \relax& 4(6)\\
14:44:47 \relax& 9(2) \relax& 143(13) \relax& 49(5) \relax& -16(10) \relax& 10(5) \relax& -12(5)\\
14:52:33 \relax& 20(3) \relax& 121(11) \relax& 39(4) \relax& -18(11) \relax& 9(6) \relax& -25(8)\\
\midrule
Average\textsuperscript{\scriptsize{a}} \relax& 11.5(0.6) \relax& 127(3) \relax& 43.0(1.2) \relax& 0(2) \relax& 22.0(1.3) \relax& 12.5(1.3)\\
\bottomrule
\multicolumn{7}{p{0.73\linewidth}}{\textbf{Notes:} We retrieved each brightness from a circular aperture with a radius of \ang{;;1.75}. Assuming emission from a disk with the solid-angular size of Ganymede (it had an apparent angular radius of \ang{;;0.775} on \mbox{2021-06-08 UTC}), we scaled the brightness by the ratio $(\ang{;;1.75}/\ang{;;0.775})^2$. Listed errors include 9\% systematic uncertainty.
\begin{itemize}[nosep, labelsep=-1.5pt, align=left, leftmargin=*]
    \item[\textsuperscript{a}]Calculated using weighted averages as \mbox{$\left<B\right> = \sum_{i=1}^{15}(B_iw_i)/\sum_{i=1}^{15}w_i$} with average uncertainty \mbox{$\left<\sigma\right>=1/\sqrt{\sum_{i=1}^{15}{w_i}}$}, where $w_i = 1/\sigma_i^2$ is the inverse variance.
\end{itemize}}
\end{tabular}
\end{table*}

We used the standard deviation of the spectrum near the emission to estimate the random error from instrumental effects and photon counting. To estimate systematic error, we compared the ratio of the \qty{630.0}{nm} brightness to the \qty{636.4}{nm} brightness to see how many of the observed ratios deviated from the expected value. The lifetime for \ce{O(^1D_2 \to ^3P_2)} \qty{630.0}{nm} emission is \qty{178}{s} while the lifetime for \ce{O(^1D_2 \to ^3P_1)} \qty{636.4}{nm} emission is \qty{549}{s} \citep{Wiese1996}, so if there is collisional quenching, the observed \qty{636.4}{nm} emission should be suppressed more than the \qty{630.0}{nm} emission and the $\qty{630.0}{nm}/\qty{636.4}{nm}$ emission ratio should be larger. However, the value cannot be any lower than the expected ratio of 3.09 for a collisionless atmosphere \citep{Wiese1996}. We found that including 9\% systematic error in addition to the random error derived from the standard deviation of the spectrum resulted in approximately two-thirds of the observed ratios to be within $1\sigma$ of the expected value. All uncertainties listed in this paper are the quadrature sum of these two errors.

\subsection{Calculation of Incident Electron Densities}

\begin{figure}
\includegraphics[width=\columnwidth]{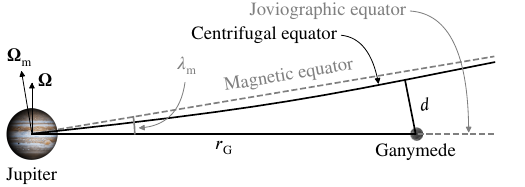}
\caption{Geometric relationships relevant to the retrieval of electron properties. Ganymede orbits Jupiter at a distance of $r_\mathrm{G}$. However, because of the tilt of the magnetic field rotation axis $\mathbf{\Omega}_\mathrm{m}$ relative to the Joviographic rotation axis $\mathbf{\Omega}$, Ganymede's distance $d$ from the centrifugal equator varies over the approximately 10 hour magnetospheric rotation period of Jupiter. See \cite{Phipps2021} for a detailed description of this geometry. Image credits: Jupiter \citepalias{JupiterGlobe}, Ganymede \citepalias{USGSMap}.}
\label{fig:plasma-sheet-geometry}
\end{figure}

The distribution of plasma in Jupiter's magnetosphere is affected by both rotational forces and particle energies. \cite{Phipps2021} showed the plasma density reaches its maximum along a centrifugal equator which exhibits a variable latitudinal offset from the magnetic field equator depending on distance from Jupiter. \cite{Bagenal2011} modeled the plasma distribution and derived how the vertical scale height of the electrons varies with radial distance from Jupiter due to competition between centrifugal forces and thermal pressure. The combination of these two effects with Jupiter's rotating magnetosphere affect the incident plasma density exciting the aurora.

To evaluate this effect quantitatively, we calculated the location of the plasma sheet mid-plane using the methodology outlined in \cite{Phipps2021}. Assuming symmetry along the azimuth axis, they derived an empirical fit to a dipole magnetic field which gives the Joviographic latitude $\lambda_\mathrm{ceq}$ of the plasma density maxima as a function of distance from Jupiter $r$ and sub-Jovian east longitude $\phi$
\begin{equation}\label{eq:latitude}
\lambda_\mathrm{ceq}(r,\phi) = \left[a\tanh\left(b\frac{r}{R_\mathrm{J}}-c\right) + d\right]\sin(\phi-e),
\end{equation}
where the empirically-derived best-fit constants are $a=\ang[angle-symbol-over-decimal=false]{1.66}$, $b=\qty{0.131}{rad}$, $c=\qty{1.62}{rad}$, $d=\ang[angle-symbol-over-decimal=false]{7.76}$ and $e=\ang{249}$.

We retrieved the distance between Jupiter and Ganymede $r_\mathrm{G}$ and the sub-Jovian west longitude $\phi_\textsc{iii,rh}$ and latitude $\phi_\textsc{iii,rh}$ in Jupiter's System III Joviographic reference frame using the JPL Horizons. We first made an initial query targeting Ganymede as observed from Maunakea and retrieved the light travel time between Ganymede and Earth. We then subtracted that travel time from the observation start time and made a second query targeting Jupiter as observed from Ganymede at the new time, retrieving ephemeris information in the Jupiter system at the time of the observation. We converted all System III coordinates to the right-handed Joviographic coordinate system.

\begin{figure}
\includegraphics[width=\columnwidth]{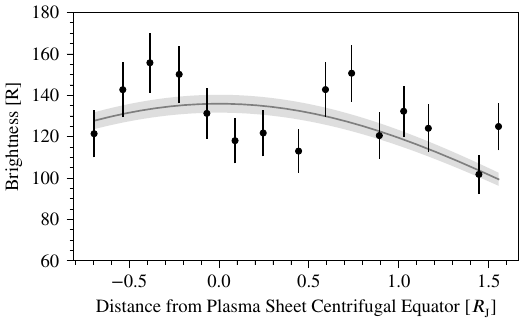}
\caption{Disk-integrated brightness of Ganymede's \qty{630.0}{nm} aurora as a function of distance from the plasma sheet mid-plane. The black points are the observations with error bars showing the combined random and systematic uncertainties. The gray line is a fit of equation (\ref{eq:exponential-brightness-scaling}) with $B_0=\qty{130(5)}{R}$ and the scale height $H$ fixed to \qty{2.78}{\RJ} \citep{Bagenal2011}. The shaded gray region shows the uncertainty in the peak brightness ($B_0$) of the fit. The fit shows that the brightness exhibits a moderate correlation with distance from the plasma sheet mid-plane, and has an expected peak brightness of about \qty{130}{R} if the oscillation is due to localized variations in plasma density. In this figure, the timing of the observations moves from right to left: Ganymede was above the plasma sheet mid-plane at the start of the observations, and moved through the center to below the sheet by the end of the night.}
\label{fig:plasma-sheet-distance}
\end{figure}

Figure \ref{fig:plasma-sheet-geometry} shows the geometric relationships relevant to the calculation of the plasma density. Ganymede orbits Jupiter at a distance $r_\mathrm{G}$, but because of the relative \ang[angle-symbol-over-decimal=false]{9.5} tilt of the magnetic field axis toward \ang{159} east longitude \citep{Connerney1998}, the height of Ganymede above the highest-density mid-plane of the plasma sheet $d$, which falls along the centrifugal equator, varies with the approximately 10-hour rotation period of Jupiter's magnetosphere. To calculate the effective plasma density $n$ incident at Ganymede, we used a mid-plane density of $n_0=\qty{20}{cm^{-3}}$ measured during the \textit{Voyager} flyby \citep{Scudder1981} scaled with a Gaussian vertical distribution of the form
\begin{equation}\label{eq:exponential-scaling}
    n = n_0 \eu^{-(d/H)^2},
\end{equation}
where $H$ is the scale height of the plasma at Ganymede's orbital distance from Jupiter $r_\mathrm{G}$ and $d$ is the minimum distance between Ganymede and the centrifugal equator, equivalent to the distance to the tangent to the centrifugal equator from which a normal line intersects Ganymede's orbit \citep{Gledhill1967}. We used a fixed scale height value of \qty{2.78}{\RJ} and orbital distance of \qty{14.936}{\RJ} \citep{Bagenal2011}. Because the auroral brightness is directly proportional to the density of the exciting electrons in the case of a thin, non-collisional atmosphere \citep{deKleer2018}, this equation also describes the expected change in brightness $B$ with distance from the centrifugal equator (see figure \ref{fig:plasma-sheet-distance}), such that it can be rewritten as
\begin{equation}\label{eq:exponential-brightness-scaling}
    B = B_0\eu^{-(d/H)^2}
\end{equation}
where $B_0$ is the peak brightness when Ganymede is at the high-density plasma sheet mid-plane.

\section{Results and Discussion}

\subsection{Disk-Integrated Brightness Variability}

The disk-integrated brightnesses of Ganymede's aurora do not correlate solely with its location relative to the plasma sheet (see figure \ref{fig:plasma-sheet-distance}). Instead, the brightness appears to exhibit a bimodal distribution, reaching local maxima when Ganymede is both above and below the mid-plane of the plasma sheet (the peaks near \qty{0.7}{\RJ} and \qty{-0.3}{\RJ}, respectively). There aren't enough high-cadence observations of Ganymede's aurora as it passes through the plasma sheet mid-plane to determine whether the dimming of the brightness between \qty{0.5}{\RJ} and the mid-plane is physically meaningful rather than either stochastic variability in local electron number density or an observational effect such as Ganymede drifting in and out of the slit. Therefore, we cannot know if this apparent bi-modality about the plasma sheet centrifugal equator is a persistent feature. However, the brightness varies smoothly in time, suggesting it isn't due to random noise. Additionally, the variability exceeds the statistical error in the individual observations. Note that time moves from right-to-left in this image; the first observations were taken when Ganymede was above the plasma sheet, and the last were taken when it was below. 

These results are less clear than similar analyses of Europa, which showed a direct correlation between distance from the plasma sheet and disk-integrated aurora brightness \citep{Roth2016,deKleer2023}. Europa's lack of a magnetic field simplifies the excitation process in comparison to Ganymede, since incident electrons at Europa are neither locally accelerated nor restricted to particular geographic locations like they are for Ganymede. However, \cite{deKleer2023} did find a potentially similar decrease in brightness when Europa was near the plasma sheet mid-plane in two sets of observations taken on \mbox{2021-06-21 UTC} and \mbox{2021-07-16 UTC}, though neither data sub-set observed Europa both above and below the mid-plane. \citet{Musacchio2017} analyzed UV observations taken by Hubble/STIS and found an increase in brightness on the leading hemisphere and a decrease in brightness on the trailing hemisphere when Ganymede was near the plasma sheet mid-plane. In contrast, the optical observations view the sub-Jovian hemisphere, and we didn't observe a change in the brightness ratio between the dusk-dawn (leading-trailing) hemispheres as Ganymede crossed through the mid-plane (see section \ref{sec:hemispheric-asymmetry} and figure \ref{fig:dusk-dawn-asymmetry}).

Io's \SI{135.6}{nm} auroral limb glow exhibits a comparable decrease in brightness with distance from the plasma sheet centrifugal equator \citep{Retherford2003}. \cite{Oliversen2001} and \cite{Schmidt2023} observed the same effect in Io's \SI{630.0}{nm} auroral emission, but the higher cadence of the optical observations revealed additional variability which \cite{Schmidt2023} attributed to heterogeneity in the local plasma density that sweeps past the satellite. Analysis of \textit{Cassini} data taken during it's flyby of Jupiter showed variations in electron number density with magnetic longitude \citep{Steffl2006,Steffl2008}. This suggests the bi-modal brightness modulation apparent in figure \ref{fig:plasma-sheet-distance} may have simply reflected local upstream plasma conditions changing over the course of the observations. Simulations of interactions between Ganymede's magnetosphere and Jupiter's also suggest that magnetic reconnection rates vary on the order of tens of seconds, affecting the supply of electrons into Ganymede's atmosphere and subsequently the number of electron-impact excitations leading to auroral emission \citep{Jia2009}.

Figure \ref{fig:plasma-sheet-distance} shows a fit of equation \eqref{eq:exponential-brightness-scaling} with the scale height $H$ fixed to a value of \qty{2.78}{\RJ} as calculated by \cite{Bagenal2011} for Ganymede's orbital distance from Jupiter. The resulting fit has a Pearson correlation coefficient of 0.415 and a \textit{p}-value of 0.124 (not achieving statistical significance for a 95\% confidence threshold). Though we expect there to be a correlation with plasma sheet distance, the poor correlation suggests two simultaneous phenomena may be affecting Ganymede's auroral brightness: first-order brightness variation from scale-height-induced density variation as its position changes relative to the plasma sheet mid-plane and second-order brightness bi-modality from localized density variations in the plasma as it sweeps past. These observations suggest the longitudinal density heterogeneity overwhelms the scale-height dependence, but further observations will be needed to confirm this conclusion statistically.

\subsection{Hemispheric Brightness Asymmetries}\label{sec:hemispheric-asymmetry}

\begin{figure*}
\includegraphics[width=\textwidth]{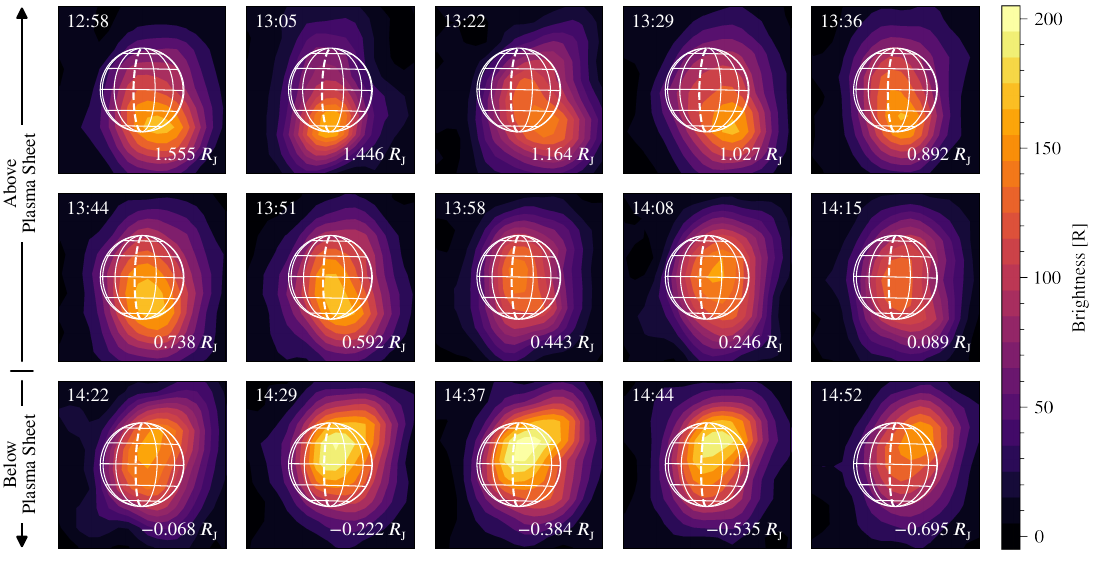}
\caption{Calibrated images of Ganymede's \qty{630.0}{nm} auroral emission displayed in \qty{20}{R} contours. To better reveal the spatial variability we smoothed the data using a Gaussian kernel with a FWHM of \ang{;;0.5} approximating typical seeing conditions for the morning of \mbox{2021-06-08 UTC}. The time in the upper left of each image is the UTC time at the start of the observation, the annotation in the lower right is the distance to the plasma sheet centrifugal equator; positive when Ganymede is above the mid-plane and negative when it is below. As Ganymede moved vertically through the plasma sheet, the hemisphere closest to the mid-plane exhibited the brightest aurora. From approximately 14:00 to \mbox{14:45 UTC}, Ganymede is within \qty{\pm 0.5}{\RJ} and the brightness is more evenly-distributed across the disk. Ganymede passed through the highest-density plasma mid-plane at \mbox{14:20 UTC}, so the top two rows display data when Ganymede was above the mid-plane (with enhanced brightness at southern latitudes) and the bottom row displays data when Ganymede was below the mid-plane (with enhanced brightness at northern latitudes). Because of the blurring effect of atmospheric seeing, the calibrated brightness distribution of the individual pixels is lower than the disk-averaged value reported in \cite{deKleer2023} and table \ref{tab:observation-information}, which assume all emission originates from a disk with Ganymede's solid-angular size. The white grid shows the physical size and orientation of Ganymede, with north pointing upward (for more details on this observing geometry, see figure \ref{fig:observing-geometry}).}
\label{fig:spatial-variation}
\end{figure*}

The \qty{630.0}{nm} emission data show both north-south and dusk-dawn (equivalent to trailing-leading in optical observations of the sub-Jovian hemisphere) hemispheric asymmetries (figure \ref{fig:spatial-variation}). 

\subsubsection{North-South Asymmetry}

When Ganymede is above the plasma sheet mid-plane (the first two rows in figure \ref{fig:spatial-variation}), the southern mid-plane-facing hemisphere is brighter. As the high-density center of the plasma sheet moves past Ganymede, the brightness is more evenly spread across the disk. Once Ganymede is below the mid-plane, the northern hemisphere is brighter. In all of the images, the peak brightness appears shifted toward dusk (trailing) hemisphere longitudes.

In order to quantify these asymmetries, we calculated the hemisphere-integrated brightnesses from semi-circular apertures with radii of \qty{2.25}{\textit{R}\textsubscript{G}}. For pixels that fell in both hemispheres we allocated the brightness based on the relative pixel area in each hemisphere. Table \ref{tab:asymmetries} lists the brightnesses of each relevant hemisphere. 

\begin{table}
\centering
\caption{[O\,\textsc{i}] \qty{630.0}{nm} brightnesses calculated for the northern, southern, dawn (leading) and dusk (trailing) hemispheres.}
\label{tab:asymmetries}
\begin{tabular}{lS[table-format=3.1(2), table-number-alignment=right]S[table-format=3.1(2), table-number-alignment=right]S[table-format=3.1(1.1), table-number-alignment=right]S[table-format=3.1(1.1), table-number-alignment=right]}
\toprule
 & \multicolumn{4}{c}{{Hemisphere-Integrated Brightness}}\\
\cmidrule(lr){2-5}
\multirow[t]{2}{*}{\shortstack[l]{Observation\\Start Time}} \relax& {North} \relax& {South} \relax& {Dawn} \relax& {Dusk}\\
{[UTC]} \relax& {[R]} \relax& {[R]} \relax& {[R]} \relax& {[R]} \\
\midrule
12:58:16 \relax& 76(8) \relax& 175(16) \relax& 88(9) \relax& 175(16)\\
13:05:29 \relax& 68(8) \relax& 136(13) \relax& 111(11) \relax& 103(10)\\
13:22:12 \relax& 83(9) \relax& 163(15) \relax& 71(8) \relax& 186(17)\\
13:29:41 \relax& 98(10) \relax& 167(16) \relax& 88(9) \relax& 187(17)\\
13:36:46 \relax& 105(11) \relax& 145(14) \relax& 106(11) \relax& 149(14)\\
13:44:33 \relax& 121(12) \relax& 183(17) \relax& 120(12) \relax& 192(18)\\
13:51:41 \relax& 116(11) \relax& 178(17) \relax& 112(11) \relax& 189(18)\\
13:58:52 \relax& 105(11) \relax& 131(13) \relax& 97(10) \relax& 142(14)\\
14:08:09 \relax& 120(12) \relax& 146(14) \relax& 87(9) \relax& 183(17)\\
14:15:28 \relax& 108(11) \relax& 135(13) \relax& 94(10) \relax& 154(15)\\
14:22:45 \relax& 149(14) \relax& 130(13) \relax& 92(10) \relax& 185(17)\\
14:29:56 \relax& 192(18) \relax& 140(13) \relax& 110(11) \relax& 220(20)\\
14:37:35 \relax& 195(19) \relax& 148(14) \relax& 105(11) \relax& 230(20)\\
14:44:47 \relax& 180(17) \relax& 128(12) \relax& 113(11) \relax& 188(18)\\
14:52:33 \relax& 152(15) \relax& 111(11) \relax& 80(9) \relax& 178(17)\\
\midrule
Average \relax& 106(3) \relax& 143(4) \relax& 95(3) \relax& 165(4)\\
\bottomrule
\end{tabular}
\end{table}

\begin{figure}
\includegraphics[width=\columnwidth]{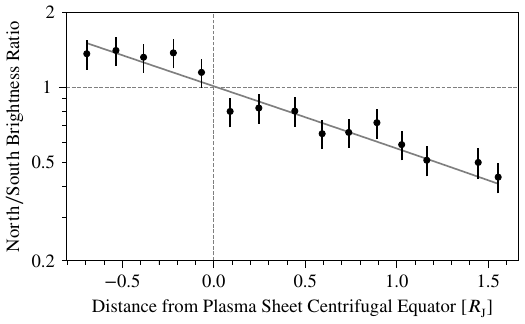}
\caption{Ratio of Ganymede's \qty{630.0}{nm} auroral emission between its northern and southern hemispheres out to an angular distance of \SI{2.25}{\textit{R}\textsubscript{G}}. The horizontal dashed gray line indicates a ratio of 1 (equal average brightness between hemispheres), and the vertical dashed gray line shows where Ganymede passed through the center of the plasma sheet. The diagonal line shows the expected brightness ratio due to asymmetric flux tube electron column densities (equation \ref{eq:ratio}).}
\label{fig:hemispheric-brightness-ratio}
\end{figure}

\cite{Retherford2003} showed that electrons in the Jovian magnetosphere impact Io's atmosphere along two different pathways: the bulk rotation of the plasma sheet which produces a flux directed at the trailing hemisphere, and bounce motion along flux tubes constrained by the morphology of Jupiter's magnetic field lines and how they connect with Io's magnetic field. They analyzed a similar north-south brightness asymmetry observed in Io's UV aurora and showed that modeled field-aligned electron motion along a flux tube accurately reproduced the brightness ratio they observed. Assuming the symmetric Gaussian profile around the centrifugal equator with a scale height of $H=\SI{2.78}{\RJ}$ (equation \ref{eq:exponential-scaling}), we numerically integrated the ratio $R_\mathrm{N/S}$ of flux tube electron column densities for a distance $d$ from the plasma sheet centrifugal equator
\begin{equation}\label{eq:ratio}
    R_\mathrm{N/S} = \frac{\displaystyle\int_d^\infty \eu^{-(x/H)^2}\,\mathrm{d}x}{\displaystyle\int_{-\infty}^d \eu^{-(x/H)^2}\,\mathrm{d}x}.
\end{equation}
Figure \ref{fig:hemispheric-brightness-ratio} shows the ratio of the northern hemisphere brightness to the southern hemisphere brightness with a logarithmic vertical axis so that the spacing of the ratios is meaningful. The diagonal gray line shows the result of the integral ratio in equation \eqref{eq:ratio} evaluated for the range of plasma sheet distances across the Ganymede observations. This line is not a fit to the data; the correlation coefficient between the expected ratio and the observations is 0.952 with $p\ll0.001$, clearly demonstrating that the asymmetric column densities along the flux tubes intersecting each hemisphere quantitatively matches the observed north-south brightness asymmetry of Ganymede's aurora.

\cite{Saur2022} looked at this same hemispheric brightness ratio in \qty{135.6}{nm} UV data taken on \mbox{2021-06-07 UTC} (the day before these HIRES observations), but the low signal-to-noise at ultraviolet wavelengths required them to integrate for longer, reducing the cadence of their time series observations. Regardless, they found a similarly-variable hemispheric brightness ratio in the ultraviolet data, and their north-south ratio varied between 1.6 and 0.4 \citep[see][figure 5]{Saur2022} which matches the extremes in the \qty{630.0}{nm} HIRES observations. At an orbital distance of \qty{14.936}{\RJ} we calculated Ganymede reaches a maximum height above or below the plasma sheet centrifugal equator of \qty{2.44}{\RJ}, so from equation \eqref{eq:ratio} the peak north-south hemispheric brightness ratio should be about 4.35 at $\qty{-2.44}{\RJ}$ and 0.230 at $\qty{2.44}{\RJ}$.

\begin{figure}
\includegraphics[width=\columnwidth]{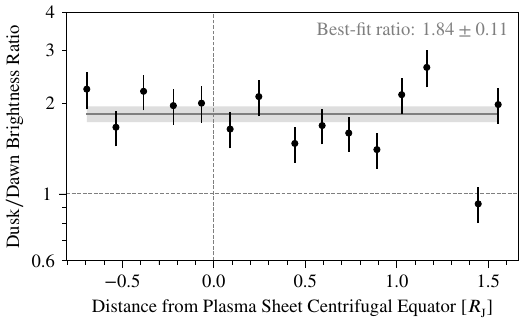}
\caption{Ratio of Ganymede's \qty{630.0}{nm} auroral emission between its dusk and dawn hemispheres out to an angular distance of \SI{2.25}{\textit{R}\textsubscript{G}}. The horizontal dashed gray line indicates a ratio of 1 (equal average brightness between hemispheres), and the vertical dashed gray line shows where Ganymede passed through the center of the plasma sheet. The horizontal line is a best-fit constant and the shaded gray area shows the uncertainty in the fit.}
\label{fig:dusk-dawn-asymmetry}
\end{figure}

\subsubsection{Dusk-Dawn Asymmetry}

Figure \ref{fig:dusk-dawn-asymmetry} shows the ratio of the dusk hemisphere brightness to the dawn hemisphere brightness. We found the dusk (trailing) hemisphere was almost always twice as bright as the dawn (leading) hemisphere. \citet{McGrath2013} observed a similar brightness asymmetry in HST observations of Ganymede's \qty{135.6}{nm} emission. \citet{Musacchio2017} and \citet{Molyneux2018} did not observe the same relative enhancement of the leading hemisphere, but they observed just the leading or trailing hemisphere (rather than the simultaneous leading-trailing geometry of the sub-Jovian optical observations), so they wouldn't be able to see the same kind of inter-hemisphere enhancement that we did.

\cite{Leblanc2017} simulated Ganymede's atmosphere and showed an enhancement in \ce{O2} column density toward the dusk hemisphere in eclipse (see their figure 5). Because \ce{O2} does not readily condense on the Ganymede's surface, they suggested the asymmetry is due to a combination of the thermal inertia of the surface ice, the morphology of the magnetic field and the incident sputtering particles originating from the direction of the trailing hemisphere. In particular, thermal lag in the surface ice causes the surface to be warmer toward dusk compared to dawn. A higher temperature on the hemisphere subjected to sputtering particle flux allows for a larger sputtering rate \citep{Cassidy2013}. \cite{Oza2018} estimated latitude-averaged \ce{O2} column densities on several tidally-locked Solar System moons, including Ganymede and Europa, and estimated a hemispherically-averaged dusk-dawn ratio of 1.22, far below our minimum estimated ratio of \num{1.84(0.11)}. The higher asymmetry found in the aurora brightnesses is likely the product of both the larger column density and a larger electron flux on the trailing hemisphere, however the optical aurora data cannot decouple their relative contributions.

Though we assumed an increased electron flux on the trailing hemisphere contributes to the dusk-dawn asymmetry, the physical structure of Ganymede's magnetosphere and the way it connects with Jupiter's almost certainly complicates the electron flux path. \citet{Kivelson2004} showed how the interaction between Jupiter's magnetosphere and a magnetized moon like Ganymede results in field line reconnection which restricts plasma flow and redirects it toward a narrow range of magnetic latitudes near the magnetic poles, allowing for electron flux on both the leading and trailing hemispheres (rather than concentrating the bulk of the flux on the trailing hemisphere). Our analysis of the north-south hemispheric brightness ratio provides evidence for flow along the field lines connecting to Jupiter and therefore access to the full vertical extent of the plasma split between the hemispheres. In contrast, simple impact on the trailing hemisphere from the rotation of Jupiter's magnetosphere would produce a much smaller ratio between the northern and southern hemispheres since the relative difference in electron flux could only come from the vertical extent of Ganymede's physical cross section. \citet{Eviatar2001} assumed an increased electron energy to account for the effects of Ganymede's magnetic field, while \citet{Saur2022} increased both the number density and the peak of the electron energy distribution \citep[to double that of][]{Eviatar2001}. Though we used the electron densities derived from \textit{Voyager} data \citep{Scudder1981} and the energy distribution given by \citet{Eviatar2001}, we evaluated the effect of the electron properties given by \citet{Saur2022} on observed optical aurora brightnesses as detailed in section \ref{sec:saur2022}.

\subsection{Constraints on Atmospheric Composition Variability over Eclipse}

\begin{table}
\centering
\caption{Modeled emission ratios relative to \qty{557.7}{nm} [\textsc{O\,i}] for Ganymede's optical aurora, assuming an electron population with a Maxwell-Boltzmann distribution centered at \qty{100}{eV} and a number density of \qty{20}{cm^{-3}}.}
\label{tab:emission-rate-coefficients}
\begin{tabular}{lS[table-format=2.4, table-number-alignment=right]S[table-format=2.4, table-number-alignment=right]S[table-format=2.4, table-number-alignment=right]S[table-format=2.4, table-number-alignment=right]}
\toprule
& \multicolumn{4}{c}{{Parent Species}}\\
\cmidrule{2-5}
 \relax& {\ce{O}} \relax& {\ce{O2}} \relax& {\ce{H2O}} \relax& {\ce{CO2}}\\
\midrule
\SI{121.6}{nm} H\,\textsc{i} \relax& {---} \relax& {---} \relax& 4.58 \relax& {---} \\
\SI{130.4}{nm} O\,\textsc{i} \relax& 18.7 \relax& 1.38 \relax& 0.178 \relax& 0.0471 \\
\SI{135.6}{nm} O\,\textsc{i}] \relax& 0.661 \relax& 3.11 \relax& 0.0416 \relax& 0.0480 \\
\SI{297.2}{nm} [O\,\textsc{i}] \relax& 0.0598 \relax& 0.0598 \relax& 0.0598 \relax& 0.0598 \\
\SI{486.1}{nm} H\,\textsc{i} \relax& {---} \relax& {---} \relax& 0.409 \relax& {---} \\
\SI{557.7}{nm} [O\,\textsc{i}] \relax& 1 \relax& 1 \relax& 1 \relax& 1 \\
\SI{630.0}{nm} [O\,\textsc{i}] \relax& 2.72 \relax& 13.5 \relax& 0.943 \relax& 0.892 \\
\SI{636.4}{nm} [O\,\textsc{i}] \relax& 0.880 \relax& 4.38 \relax& 0.305 \relax& 0.288 \\
\SI{656.3}{nm} H\,\textsc{i} \relax& {---} \relax& {---} \relax& 2.27 \relax& {---} \\
\SI{777.4}{nm} O\,\textsc{i} \relax& 0.241 \relax& 2.06 \relax& 0.0716 \relax& 0.0282 \\
\SI{844.6}{nm} O\,\textsc{i} \relax& 3.24 \relax& 0.972 \relax& 0.186 \relax& 0.0205\\
\bottomrule
\end{tabular}
\end{table}

The ratio of the \qty{630.0}{nm} brightness to the \qty{557.7}{nm} brightness is particularly sensitive to the presence of \ce{H2O} as a parent molecule \citep[table \ref{tab:emission-rate-coefficients}, see also][table 5]{deKleer2023}. Since the publication of \cite{deKleer2023} we have expanded the aurora model to include cross-sections for electron impact on \ce{CO2} producing emission at \qty{630.0}{nm} and \qty{636.4}{nm} \citep{Strickland1969} and \qtylist{777.4;844.6}{nm} \citep{Zipf1984}. Using the same cross sections for \ce{O}, \ce{O2} and \ce{H2O} listed in \citet{deKleer2023}, we've calculated a \qty{630.0}{nm} to \qty{557.7}{nm} emission ratio of $13.5$ for electron impact on \ce{O2}, $2.72$ for for electron impact on \ce{O}, $0.943$ for for electron impact on \ce{H2O} and 0.892 for electron impact on \ce{CO2}. The \ce{H2O} ratio is nearly 1, so we should observe approximately equal brightnesses at 557.7 and \qty{630.0}{nm} if the primary source of the auroral emission was electron impact on water molecules.

\begin{figure}
\includegraphics[width=\columnwidth]{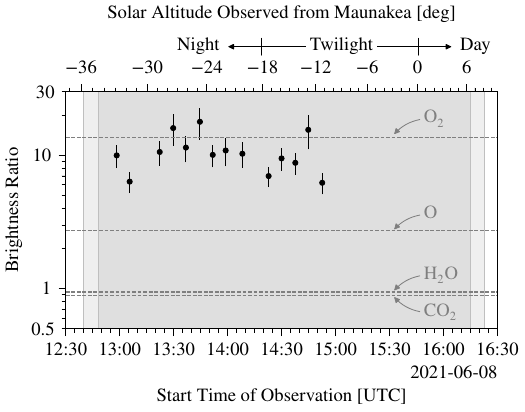}
\caption{Time series of the ratio of the aurora brightness at \qty{630.0}{nm} to the brightness at \qty{557.7}{nm}. We have selected only data points for which the ratio signal-to-noise was greater than 2, which eliminated the observation at 14:15:28. Dashed horizontal lines show the modeled ratios for atmospheres of pure O, \ce{O2}, \ce{H2O} and \ce{CO2}}. The ratios remain relatively constant over the duration of the observed eclipse and do not show evidence for the presence of significant \ce{H2O} or \ce{CO2}. The dark gray background region spanning most of the time range shows the 3-hour, 47-minute duration of the umbral eclipse. The light gray regions on either side are the approximately 8-minute duration of the partial umbral eclipse. The top axis shows the altitude of the Sun as observer from the summit of Maunakea. Sunrise on Maunakea occurred at \mbox{15:46 UTC}, preventing observations of the end of the eclipse.
\label{fig:ratio-time-series}
\end{figure}

Figure \ref{fig:ratio-time-series} shows a time-series of the \qty{630.0}{nm} to \qty{557.7}{nm} emission ratio retrieved from the individual Ganymede observations. The ratio appears relatively constant over the duration of the eclipse, though on average lower than expected for a pure-\ce{O2} atmosphere, suggesting the presence of an atmospheric species beyond \ce{O2}. If Ganymede's atmosphere is collisional, quenching of the longer-lived \ce{O(^1D_2)} atoms \citep{Wiese1996} could also lower the ratio of \qty{630.0}{nm} to \qty{557.7}{nm} emission. (Transitions from \ce{O(^1D_2)} emit both the \qtylist{630.0;636.4}{nm} photons when relaxing to the \ce{O(^3P_2)} and \ce{O(^3P_1)} ground states, respectively.) However, we did not find any evidence of collisional quenching in our analysis of the $\qty{630.0}{nm}/\qty{636.4}{nm}$ emission ratio when estimating systematic error (see section \ref{sec:systematics}).

The relatively constant ratio suggests atmospheric composition was not changing over the course of the observations; if there was an \ce{H2O} atmosphere, it either collapsed within the first ten minutes of umbral eclipse (an effect we analyzed in section \ref{sec:detectability}) or the emissions we observed included only a minor contribution from \ce{H2O}.

\begin{figure}
\includegraphics[width=\columnwidth]{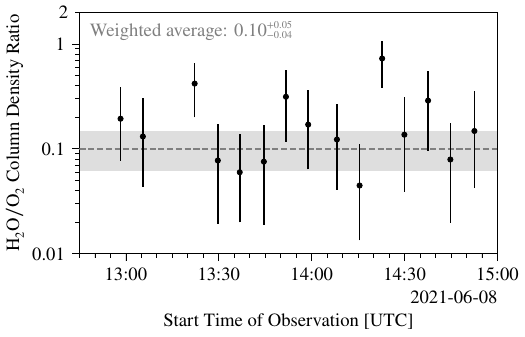}
\caption{$\ce{H2O}/\ce{O2}$ column density ratio derived from MCMC fits to retrieved aurora brightnesses. The dashed gray line shows the weighted average of the individual points and the shaded gray region shows the asymmetric uncertainty of the weighted average. This average column density ratio of $0.10${\raisebox{0.5ex}{\tiny$^{+0.05}_{-0.04}$}} matches the upper limit found by \cite{deKleer2023} for the same date.}
\label{fig:mixing-ratio-time-series}
\end{figure}

We used the same Markov chain Monte Carlo (MCMC) method described in \cite{deKleer2023} to estimate the relative column densities of a four-species atmosphere consisting of \ce{O}, \ce{O2}, \ce{H2O} and \ce{CO2} using an electron number density of \qty{20}{cm^{-3}} \citep{Scudder1981} and a Maxwell-Boltzmann velocity distribution centered at \qty{100}{eV} \citep{Eviatar2001}. Table \ref{tab:atmospheres} lists the median atmospheres for each observation along with the median atmosphere to the average brightnesses listed in table \ref{tab:observation-information}.

\begin{table}
\centering
\caption{MCMC median atmospheres using retrieved brightnesses listed in table \ref{tab:observation-information}. The uncertainties listed are the 16th and 84th-percentile quantiles.}
\label{tab:atmospheres}
\begin{tabular}{lllll}
\toprule
& \multicolumn{4}{c}{Column Density [$\times 10^{18}$]}\\
\cmidrule(lr){2-5}
\multirow[t]{2}{*}{\shortstack[l]{Observation\\Start Time}} \relax& \multicolumn{1}{c}{\ce{O}} \relax& \multicolumn{1}{c}{\ce{O2}} \relax& \multicolumn{1}{c}{\ce{H2O}} \relax& \multicolumn{1}{c}{\ce{CO2}}\\
\relax [UTC] \relax& \multicolumn{1}{c}{[\si{m^{-2}}]} \relax& \multicolumn{1}{c}{[\si{m^{-2}}]} \relax& \multicolumn{1}{c}{[\si{m^{-2}}]} \relax& \multicolumn{1}{c}{[\si{m^{-2}}]}\\
\midrule
12:58:16 \relax& $0.08${\raisebox{0.5ex}{\tiny$^{+0.08}_{-0.06}$}} \relax& $5.17${\raisebox{0.5ex}{\tiny$^{+0.17}_{-0.17}$}} \relax& $1.0${\raisebox{0.5ex}{\tiny$^{+1.0}_{-0.6}$}} \relax& $0.08${\raisebox{0.5ex}{\tiny$^{+0.09}_{-0.05}$}}\\
13:05:29 \relax& $0.19${\raisebox{0.5ex}{\tiny$^{+0.12}_{-0.11}$}} \relax& $4.57${\raisebox{0.5ex}{\tiny$^{+0.17}_{-0.16}$}} \relax& $0.6${\raisebox{0.5ex}{\tiny$^{+0.7}_{-0.4}$}} \relax& $0.24${\raisebox{0.5ex}{\tiny$^{+0.14}_{-0.13}$}}\\
13:22:12 \relax& $0.09${\raisebox{0.5ex}{\tiny$^{+0.09}_{-0.06}$}} \relax& $5.01${\raisebox{0.5ex}{\tiny$^{+0.17}_{-0.17}$}} \relax& $2.0${\raisebox{0.5ex}{\tiny$^{+1.3}_{-1.1}$}} \relax& $0.06${\raisebox{0.5ex}{\tiny$^{+0.08}_{-0.04}$}}\\
13:29:41 \relax& $0.09${\raisebox{0.5ex}{\tiny$^{+0.09}_{-0.06}$}} \relax& $5.17${\raisebox{0.5ex}{\tiny$^{+0.17}_{-0.17}$}} \relax& $0.4${\raisebox{0.5ex}{\tiny$^{+0.5}_{-0.3}$}} \relax& $0.04${\raisebox{0.5ex}{\tiny$^{+0.05}_{-0.03}$}}\\
13:36:46 \relax& $0.11${\raisebox{0.5ex}{\tiny$^{+0.10}_{-0.07}$}} \relax& $5.01${\raisebox{0.5ex}{\tiny$^{+0.16}_{-0.16}$}} \relax& $0.4${\raisebox{0.5ex}{\tiny$^{+0.5}_{-0.2}$}} \relax& $0.06${\raisebox{0.5ex}{\tiny$^{+0.07}_{-0.04}$}}\\
13:44:33 \relax& $0.07${\raisebox{0.5ex}{\tiny$^{+0.08}_{-0.05}$}} \relax& $5.29${\raisebox{0.5ex}{\tiny$^{+0.16}_{-0.16}$}} \relax& $0.4${\raisebox{0.5ex}{\tiny$^{+0.5}_{-0.3}$}} \relax& $0.04${\raisebox{0.5ex}{\tiny$^{+0.05}_{-0.03}$}}\\
13:51:41 \relax& $0.09${\raisebox{0.5ex}{\tiny$^{+0.09}_{-0.06}$}} \relax& $5.08${\raisebox{0.5ex}{\tiny$^{+0.17}_{-0.17}$}} \relax& $1.5${\raisebox{0.5ex}{\tiny$^{+1.2}_{-1.0}$}} \relax& $0.12${\raisebox{0.5ex}{\tiny$^{+0.12}_{-0.08}$}}\\
13:58:52 \relax& $0.18${\raisebox{0.5ex}{\tiny$^{+0.12}_{-0.10}$}} \relax& $4.68${\raisebox{0.5ex}{\tiny$^{+0.16}_{-0.17}$}} \relax& $0.7${\raisebox{0.5ex}{\tiny$^{+0.8}_{-0.5}$}} \relax& $0.05${\raisebox{0.5ex}{\tiny$^{+0.07}_{-0.04}$}}\\
14:08:09 \relax& $0.14${\raisebox{0.5ex}{\tiny$^{+0.11}_{-0.08}$}} \relax& $4.88${\raisebox{0.5ex}{\tiny$^{+0.16}_{-0.17}$}} \relax& $0.6${\raisebox{0.5ex}{\tiny$^{+0.8}_{-0.4}$}} \relax& $0.09${\raisebox{0.5ex}{\tiny$^{+0.10}_{-0.06}$}}\\
14:15:28 \relax& $0.27${\raisebox{0.5ex}{\tiny$^{+0.13}_{-0.12}$}} \relax& $4.45${\raisebox{0.5ex}{\tiny$^{+0.16}_{-0.17}$}} \relax& $0.2${\raisebox{0.5ex}{\tiny$^{+0.3}_{-0.14}$}} \relax& $0.019${\raisebox{0.5ex}{\tiny$^{+0.03}_{-0.013}$}}\\
14:22:45 \relax& $0.09${\raisebox{0.5ex}{\tiny$^{+0.09}_{-0.06}$}} \relax& $4.93${\raisebox{0.5ex}{\tiny$^{+0.17}_{-0.17}$}} \relax& $3.7${\raisebox{0.5ex}{\tiny$^{+1.6}_{-1.7}$}} \relax& $0.17${\raisebox{0.5ex}{\tiny$^{+0.16}_{-0.12}$}}\\
14:29:56 \relax& $0.09${\raisebox{0.5ex}{\tiny$^{+0.09}_{-0.06}$}} \relax& $5.11${\raisebox{0.5ex}{\tiny$^{+0.17}_{-0.18}$}} \relax& $0.7${\raisebox{0.5ex}{\tiny$^{+0.9}_{-0.5}$}} \relax& $0.23${\raisebox{0.5ex}{\tiny$^{+0.14}_{-0.13}$}}\\
14:37:35 \relax& $0.07${\raisebox{0.5ex}{\tiny$^{+0.08}_{-0.04}$}} \relax& $5.19${\raisebox{0.5ex}{\tiny$^{+0.17}_{-0.18}$}} \relax& $1.6${\raisebox{0.5ex}{\tiny$^{+1.4}_{-1.0}$}} \relax& $0.27${\raisebox{0.5ex}{\tiny$^{+0.17}_{-0.15}$}}\\
14:44:47 \relax& $0.1${\raisebox{0.5ex}{\tiny$^{+0.10}_{-0.07}$}} \relax& $5.04${\raisebox{0.5ex}{\tiny$^{+0.17}_{-0.17}$}} \relax& $0.3${\raisebox{0.5ex}{\tiny$^{+0.5}_{-0.2}$}} \relax& $0.06${\raisebox{0.5ex}{\tiny$^{+0.07}_{-0.04}$}}\\
14:52:33 \relax& $0.15${\raisebox{0.5ex}{\tiny$^{+0.12}_{-0.09}$}} \relax& $4.69${\raisebox{0.5ex}{\tiny$^{+0.18}_{-0.18}$}} \relax& $0.8${\raisebox{0.5ex}{\tiny$^{+1.0}_{-0.5}$}} \relax& $0.45${\raisebox{0.5ex}{\tiny$^{+0.16}_{-0.17}$}}\\
\midrule
Average \relax& $0.11${\raisebox{0.5ex}{\tiny$^{+0.09}_{-0.07}$}} \relax& $5.13${\raisebox{0.5ex}{\tiny$^{+0.15}_{-0.15}$}} \relax& $0.4${\raisebox{0.5ex}{\tiny$^{+0.4}_{-0.3}$}} \relax& $0.07${\raisebox{0.5ex}{\tiny$^{+0.06}_{-0.05}$}}
\\
\bottomrule
\end{tabular}
\end{table}

For each individual observation, the median atmospheres tended toward a primarily-\ce{O2} composition, but included minor contributions from both \ce{O} and \ce{H2O} to account for the brightness ratio of less than 13.5 expected for a pure \ce{O2} atmosphere. \ce{CO2} is a trace component of Ganymede's surface ice \citep{McCord1998}, and our model found a correspondingly minor \ce{CO2} component about one order of magnitude lower than \ce{H2O} for each observation. The brightness contribution from a \ce{CO2} column of this magnitude is less than the measurement uncertainties of the auroral brightnesses, so we treat this result as an upper limit on a potential \ce{CO2} component in Ganymede's atmospheric composition.

Figure \ref{fig:mixing-ratio-time-series} shows the modeled $\ce{H2O}/\ce{O2}$ column density ratio for each observation. The weighted average (shown in gray) is $0.10${\raisebox{0.5ex}{\tiny$^{+0.05}_{-0.04}$}}. This ratio matches the $2\sigma$ upper limit of 0.10 found by \cite{deKleer2023} for the \mbox{2021-06-08 UTC} observations using averages of the spectra and still suggests an \ce{O2}-dominated atmosphere. (Though \cite{deKleer2023} report an upper limit of 0.06 for the $\ce{H2O}/\ce{O2}$ column ratio, they calculated this value using averages over multiple nights of Ganymede observations; their best-fit atmosphere for the \mbox{2021-06-08 UTC} data had an upper limit ratio of 0.10.) 

\subsection{Detectability of an H\textsubscript{2}O Atmosphere with Keck/HIRES}\label{sec:detectability}


\begin{figure*}
\includegraphics[width=\textwidth]{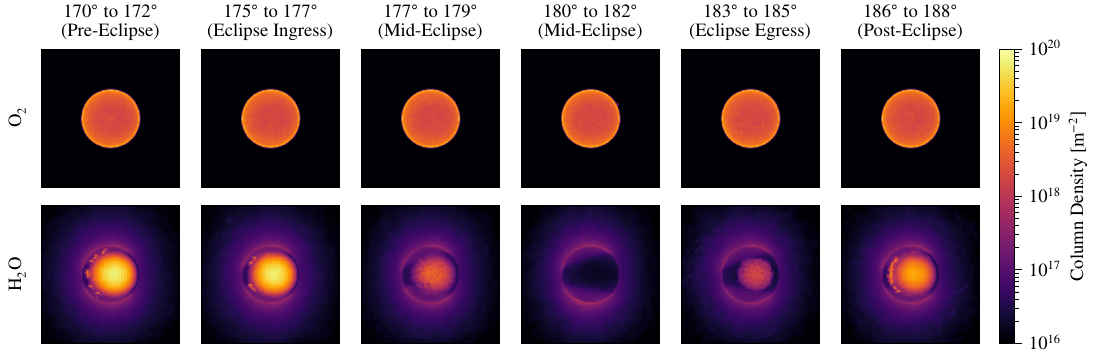}
\caption{Simulation of Ganymede's \ce{O2} (top row) and \ce{H2O} (bottom row) atmospheric column densities before, during and after eclipse by Jupiter as observed from Earth. Each time step was averaged over $2^\circ$ (57 minutes) of Ganymede's orbit. The coordinate system is right-handed, measured counter-clockwise along Ganymede's orbit with \ang{0} pointing toward the Sun and \ang{180} at the mid-point of the eclipse. Small-scale structures prominent in the \ce{H2O} column densities are artifacts from the Monte Carlo simulation method. These simulations suggest the \ce{H2O} atmosphere, if present, collapses rapidly after the onset of the eclipse, and recovers quickly after Ganymede emerges back into sunlight.}
\label{fig:sim-images}
\end{figure*}

\cite{deKleer2023} found that the column density of \ce{H2O} derived by \cite{Roth2021} would have produced more emission at \qtylist{557.7;656.3}{nm} than they detected in the average disk-integrated optical observations. However, because \ce{O2} and \ce{H2O} may have different spatial distributions \citep{Roth2021,Leblanc2017,Leblanc2023}, the observing geometry during eclipse could reduce observed emission from electron impact on \ce{H2O} due to the physical separation between the open/closed field line boundary and the higher-density \ce{H2O} column near the disk center. In order to explore this spatial effect on the aurora emission, we simulated optical observations assuming the modeled atmosphere from \cite{Leblanc2023}, which was tuned to reproduce the \cite{Roth2021} UV observations. 

\begin{figure}
\includegraphics[width=\columnwidth]{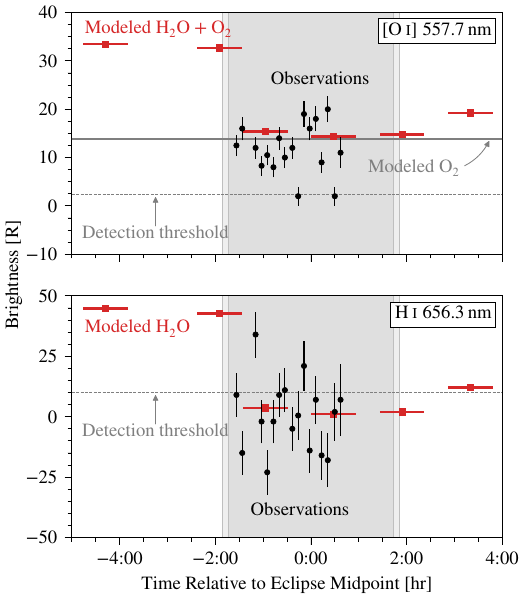}
\caption{Comparison between \qty{557.7}{nm}~[\textsc{O\,i}] and \qty{656.3}{nm}~\textsc{H\,i} (H-$\mathrm{\alpha}$) disk-integrated brightness observations and predicted brightnesses from modeled column densities. Black circles with vertical error bars show observations and their associated uncertainties. Red squares show predicted brightnesses from column densities averaged over 57 minutes (\ang{2}) of Ganymede's orbit (the horizontal bars show the extent of the averaging window). The dashed gray horizontal line in each plot shows the detection threshold (the typical standard deviation of the observations). The \qty{557.7}{nm}~[\textsc{O\,i}] modeled brightness from the \ce{O2} atmosphere does not substantially change over the course of the eclipse, so we have shown its value as a solid horizontal dark gray line. The shaded gray regions in the background are the same eclipse boundaries shown in figure~\ref{fig:ratio-time-series}. For both the \qty{557.7}{nm}~[\textsc{O\,i}] and \qty{656.3}{nm}~\textsc{H\,i} observed brightnesses, the \ce{H2O} contribution quickly drops below the detection threshold. Because the first two observations should exhibit some contribution from \ce{H2O} above the detection threshold but do not appear significantly brighter than the rest of the eclipse observations, we concluded there was no emission from a localized \ce{H2O} atmosphere present in this data set.}
\label{fig:sim-data-comparison}
\end{figure}

\begin{figure}
\includegraphics[width=\columnwidth]{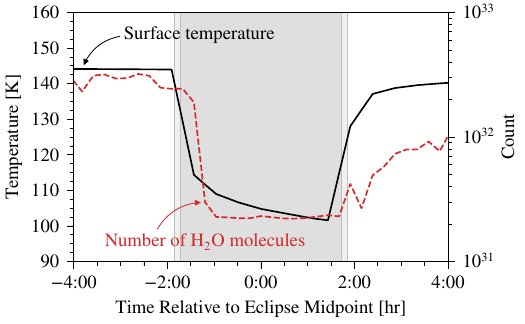}
\caption{Simulated sub-solar surface temperature \citep{Leblanc2017} and number of simulated \ce{H2O} molecules before, during and after eclipse. Ganymede's surface temperature dropped rapidly at the onset of eclipse, decreasing by \SI{30}{K} in the first half hour and a further \SI{5}{K} by the end of the first hour. The number of simulated \ce{H2O} molecules in the simulation (proportional to the \ce{H2O} number density) decreased by an order of magnitude within the first hour, and remained relatively constant for the rest of the eclipse even though the temperature continued to decrease by an additional \SI{7}{K} before the end of the eclipse.}
\label{fig:sim-temperature-density}
\end{figure}

We evaluated simulations of both \ce{O2} and \ce{H2O} atmospheres (figure \ref{fig:sim-images}). Each time step was averaged over \ang{2} of Ganymede's orbit around Jupiter (a duration of 57 minutes). To simulate eclipse conditions, the effects of solar photon ionization and dissociation were turned off between orbital angles of \ang[angle-symbol-over-decimal=false]{176.26} and \ang[angle-symbol-over-decimal=false]{183.74}, measured in a right-handed coordinate system from a reference position of \ang{0} pointing toward the Sun. These angles correspond approximately to the midpoint of the partial umbral eclipses. In our analysis of the simulations (figures \ref{fig:sim-data-comparison} and \ref{fig:sim-temperature-density}), we converted this angular coordinate system to time relative to the eclipse midpoint.

The simulated \ce{O2} atmospheres have a disk-integrated column density between \SIlist{4.45e18;4.5e18}{m^{-2}}, which are between 5 and 13\% lower than the average column density from our best-fit model atmosphere and that of \citet{deKleer2023}. The simulated \ce{H2O} disk-integrated column densities vary between about \SI{1e19}{m^{-2}} before eclipse ingress to a minimum of about \SI{7e16}{m^{-2}} just before eclipse egress. The peak simulated \ce{H2O} column density decreases from \SI{6e19}{m^{-2}} to \SI{2e16}{m^{-2}}.

We used the same Maxwell-Boltzmann electron population centered at \qty{100}{eV} as described in \cite{Leblanc2023}, with the density modified spatially to account for acceleration near the open/closed field line boundaries. We set the background number density to \qty{20}{cm^{-3}}, increasing to a peak of \qty{70}{cm^{-3}} at the latitudes of the ovals from \cite{Duling2022} following a Gaussian shape with a FWHM of \ang{20}, chosen by \cite{Leblanc2023} to match simulations of Ganymede's magnetosphere \citep{Jia2009}.

We calculated the auroral emission at the native resolution of the simulations, then rebinned the results to the detector resolution of HIRES. Finally, we smoothed the data using a two-dimensional Gaussian kernel with a FWHM of \ang{;;0.5}, representative of the typical seeing conditions of the night of \mbox{2021-06-08 UTC}. We then calculated the disk-integrated brightnesses with the same aperture size we used for the HIRES observations.

Figure \ref{fig:sim-data-comparison} shows a comparison between the simulated and observed brightnesses for \qty{557.7}{nm}~[\textsc{O\,i}] and \qty{656.3}{nm}~\textsc{H\,i} auroral emission. The six time steps in the simulated brightnesses shown in each plot in red correspond to the total emission from both the \ce{H2O} and \ce{O2} simulations. The simulated column density of \ce{O2} doesn't change substantially over the course of the eclipse, maintaining a steady simulated disk-integrated \qty{557.7}{nm} brightness of about \qty{13}{R}. The simulated steady-state \ce{H2O} atmosphere in full sunlight has a disk-integrated column density ratio with the \ce{O2} atmosphere of about 3, and our aurora model produces a simulated \qty{20}{R} of emission at \qty{557.7}{nm} and \qty{45}{R} of emission at \qty{656.3}{nm} from the model \ce{H2O} atmosphere alone. Combining the model \ce{H2O} and \ce{O2} atmospheres, the simulated \qty{557.7}{nm} brightness reaches \qty{33}{R}, almost triple the average observed value of \SI{11.5(0.6)}{R}. \cite{deKleer2023} showed Ganymede's disk-integrated aurora were extremely consistent across three different nights of observation spanning more than 20 years, so random variability in the incident electron densities alone likely cannot account for the difference between the modeled sunlit \ce{H2O} and \ce{O2} brightness and the observed brightness.

A possible explanation for the apparent lack of emission from electron impact on \ce{H2O} is the rapid condensation of the \ce{H2O} atmosphere onto the surface at the onset of eclipse, an effect seen in the simulated \ce{H2O} atmosphere. The simulated column density of \ce{H2O} decreases substantially between the second simulation time step (which includes the onset of eclipse) and the third simulation time step (fully in eclipse). The modeled sunlit \qtylist{557.7;656.3}{nm} emissions from electron impact on \ce{H2O} are well above our detection threshold, whereas the modeled in-eclipse emissions are near or below our detection threshold. (The plotted detection thresholds are the average of the combined random and systematic uncertainties and represent the typical noise level at a given wavelength.) This suggests we would be able to detect the presence of an actively-condensing \ce{H2O} atmosphere as modeled by \cite{Leblanc2023} with observations taken sufficiently early during the eclipse. Instead, the observed emissions show no temporal changes in either the 557.7 or \qty{656.3}{nm} brightnesses over the first tens of minutes of eclipse (figure \ref{fig:sim-data-comparison}). If there was a localized sublimation \ce{H2O} atmosphere present in sunlight that froze out in eclipse, the deposition process must occur within the first ten to fifteen minutes of Ganymede entering the full umbral eclipse.

Figure \ref{fig:sim-temperature-density} shows the simulated surface temperature of Ganymede's sub-solar point before, during and after eclipse \citep{Leblanc2017} along with the number of \ce{H2O} molecules in the Monte Carlo simulation. The model only tracks molecules in gas phase, so as the \ce{H2O} molecules freeze onto the surface, the number of simulated particles proportionally decreases. The simulated surface temperature of Ganymede decreased from \SI{143}{K} to \SI{113}{K} at a rate of \SI{-1}{K.min^{-1}} during the first 30 minutes of the eclipse. By the end of the eclipse the surface temperature had decreased by a further \SI{12}{K} to a minimum of \SI{101}{K}. Within the first hour, the number of simulated \ce{H2O} molecules (a quantity proportional to \ce{H2O} density) decreased by an order of magnitude.

We evaluated the change in \ce{H2O} column density over the duration of the first three observations to see what emission we could expect to detect from the collapsing \ce{H2O} atmosphere in the simulation. During the time interval that corresponds to the first HIRES observation (integrated from 10 to 15 minutes after the start of the full umbral eclipse), the model temperature fell from \SI{122}{K} to \SI{117}{K} and the column density decreased from 73\% to 66\% of the pre-eclipse maximum. Auroral brightness is directly proportional to column density, so at \SI{557.7}{nm} we would expect to measure the \SI{13.5}{R} of constant emission from electron impact on \ce{O2} and an equivalent brightness from electron impact on \ce{H2O}, for a total of about \SI{27}{R}, almost $8\sigma$ above the observed brightness of \SI{12(2)}{R}. At \SI{656.3}{nm} we would expect to measure \SI{27.6}{R} of emission from electron impact on \ce{H2O}, about $2\sigma$ above the observed brightness of \SI{9(9)}{R}.

By the start of the time interval covered by the second observation (integrated from 17 to 22 minutes after the start of the full umbral eclipse), the simulated temperature decrease has slowed and the surface temperature only changed by \SI{1}{K} from \SI{114}{K} to \SI{113}{K}. The column density decreased from 63\% to 45\% of the pre-eclipse maximum, so from electron impact on \ce{H2O} we would expect \SI{10.5}{R} of emission at \SI{557.7}{nm} for a total of \SI{24}{R} ($5\sigma$ above the observed brightness) and \SI{21.6}{R} of emission at \SI{656.3}{nm} ($2.4\sigma$ above the noise threshold).

At the start of the time interval covered by the third observation (integrated from 34 minutes to 39 minutes after the start of the full umbral eclipse), the surface temperature has dropped to about \SI{111}{K} and the column density has decreased to 9\% of the pre-eclipse maximum. The contribution to the brightnesses at both \SIlist{557.7;656.3}{nm} from electron impact on \ce{H2O} are below the noise level of the individual observations.

These simulations suggest we should detect a contribution from electron impact on \ce{H2O} to the observed brightnesses at \SIlist{557.7;656.3}{nm} in the first two observations. However, for both observations the observed brightnesses are consistent with the presence of at most a minor \ce{H2O} atmospheric column.

\subsection{Potential Impact of New Juno-Derived Electron Properties on Aurora Interpretation}\label{sec:saur2022}

\cite{Saur2022} provide an important caveat to the interpretation of modeled aurora brightnesses for Ganymede. Because of local electron acceleration within Ganymede's magnetic field, the upstream electron distribution used in most models (including ours) does not accurately represent the electron distribution that produces the aurora, and the \qty{70}{cm^{-3}} density enhancement near the auroral ovals may not be physically accurate. \cite{Greathouse2022} reported Juno observations of the \qtylist{130.4;135.6}{nm} UV aurora at very high spatial resolution. They showed the emission was narrowly-confined in latitude near the open/closed field line boundary, with combined brightnesses peaking around \qty{1000}{R}. \cite{Saur2022} evaluated the electron distribution necessary to excite aurora of these combined brightnesses and found the electrons must have a Maxwell-Boltzmann distribution centered at \qty{200}{eV} (twice the assumed energy of the upstream electrons) and a density of \qty{950}{cm^{-3}} (about 50-times higher than the assumed upstream electron number density). The significantly smaller brightnesses found by our work and previous UV studies are due to coarse spatial resolution spreading the brightness over resolution elements much larger than the emitting region.

\begin{figure}
\includegraphics[width=\columnwidth]{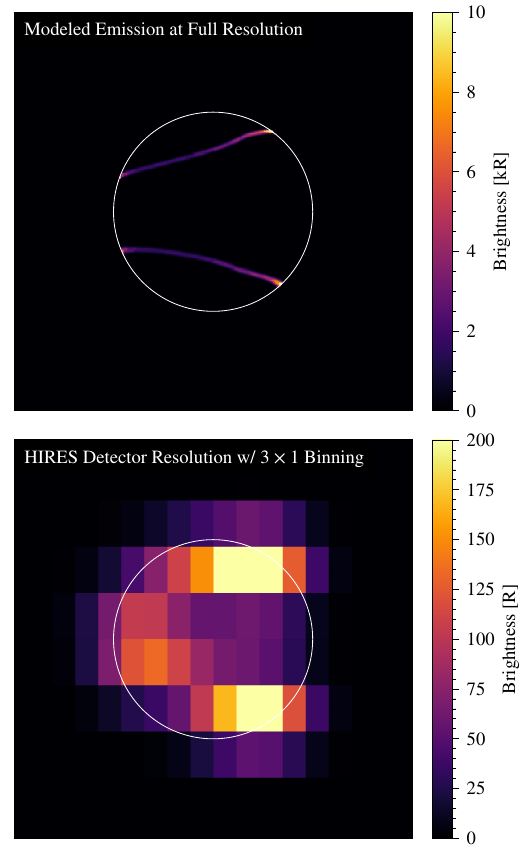}
\caption{Simulation of \qty{630.0}{nm} auroral emission from electron impact on \ce{O2} incorporating an updated electron energy distribution and emitting region. We used a Maxwell-Boltzmann electron energy distribution centered at \qty{200}{eV} with a number density of \qty{950}{cm^{-3}} \citep{Saur2022}, a typical \ce{O2} column density during eclipse and we restricted the aurora emission to a Gaussian shape along the open/closed field line boundaries with a FWHM of \ang{5} \citep{Greathouse2022}. The top high-resolution image shows the spatially-confined emission from discrete elements \ang{;;0.01} wide, with brightnesses peaking above \qty{10}{kR} at the limb. The bottom image shows this same image scaled down to the detector resolution and binning used in the Ganymede eclipse observations on \mbox{2021-06-08 UTC}. We convolved this image with a two-dimensional Gaussian kernel with a FWHM of \ang{;;0.5}, approximating typical seeing conditions on the summit of Maunakea. The white circle shows the size of the apparent disk of Ganymede. This simulation has equal electron flux for the northern and southern hemispheres, and is therefore representative of the plasma conditions encountered by Ganymede at the plasma sheet mid-plane.}
\label{fig:sim-saur2022-params}
\end{figure}

To explore the manifestation of this effect in the HIRES observations, we evaluated the \qty{630.0}{nm} emission using these electron properties and the simulated \ce{O2} column density from \ang{177} to \ang{179} (see figure \ref{fig:sim-images}). We restricted the emission to the latitudes of the open/closed field line boundaries \citep{Duling2022} using a Gaussian profile with a FWHM of \ang{5} \citep{Greathouse2022}. To account for the leading-trailing brightness asymmetries apparent in the HIRES observations, we applied a sinusoidal scaling to the electron densities with longitude, decreasing from \qty{950}{cm^{-3}} at the trailing hemisphere limb longitude to 26\% of this value (\qty{244}{cm^{-3}}) at the leading hemisphere limb longitude. This produces an electron flux with a dusk-dawn asymmetry matching the value of 1.84 we found in our analysis of the HIRES observations. Though we assume this brightness gradient is due exclusively to spatially-varying electron number density, it could also incorporate a higher \ce{O2} column density on the trailing hemisphere \citep{Oza2018}.

Figure \ref{fig:sim-saur2022-params} shows the simulated physical emission from Ganymede with these parameters at full-resolution in the top image and at HIRES detector resolution in the bottom image. Both of these simulations use the modeled \ce{O2} column densities shown in figure \ref{fig:sim-images}. The total simulated disk-integrated brightness is \qty{149}{R}, which we retrieved from the simulation using the same aperture size as the data. This result is extremely close to the best-fit value of \qty{130(5)}{R} at \qty{630.0}{nm} retrieved from the HIRES observations for the center of the plasma sheet (figure \ref{fig:plasma-sheet-distance}), especially considering the simplistic assumptions of this simulation; small changes in the electron gradient could easily produce a result matching the observed brightnesses. As a result, we concluded that the observations are therefore consistent with the electron properties derived by \cite{Saur2022} when coupled with the emission spatial distribution and hemispheric brightness asymmetry from \cite{Greathouse2022} and the pixel scale of the HIRES detectors. 

This further demonstrates that either choice of electron energy distribution, number density and spatial distribution is consistent with the derived column densities of this and other works. Table \ref{tab:emission-rate-coefficients-hi-energy} lists the expected emission ratios for this higher energy and higher number density Juno-derived electron distribution. The \qty{200}{eV} electron energy distribution from \citet{Saur2022} is consistent with the \textit{in situ} measurements of the low-energy electron distribution reported by \citet{Allegrini2022}, which were made from within Ganymede's magnetosphere by the Jovian Auroral Distributions Experiment (JADE) during the Juno flyby of Ganymede on 2021-06-07 UTC. They did not report any corresponding measurements of electron number density from within Ganymede's magnetosphere, but outside of it they found densities that were typically between \qtylist{5;20}{cm^{-3}}, more consistent with the \textit{Voyager} flyby results \citep{Scudder1981}. Similarly, \citet{Kurth2022} used measurements of total electron densities made by the Juno/Waves instrument during the same Ganymede flyby and found they varied between \SIrange{15}{30}{cm^{-3}}. However, when we simulated auroral emission like that shown in figure \ref{fig:sim-saur2022-params} assuming the same column densities but using with a Maxwell-Boltzmann electron energy distribution centered at \qty{200}{eV} and a number density of \qty{20}{cm^{-3}}, we found a predicted emission of only \qty{3}{R} of \qty{630.0}{nm}, suggesting more complex acceleration processes must be occurring within Ganymede's magnetic field.

\begin{table}
\centering
\caption{Modeled emission ratios relative to \qty{557.7}{nm} [\textsc{O\,i}] for Ganymede's optical aurora, assuming the Juno-derived electron population with a Maxwell-Boltzmann distribution centered at \qty{200}{eV} and a number density of \qty{950}{cm^{-3}}.}
\label{tab:emission-rate-coefficients-hi-energy}
\begin{tabular}{lS[table-format=2.4, table-number-alignment=right]S[table-format=2.4, table-number-alignment=right]S[table-format=2.4, table-number-alignment=right]S[table-format=2.4, table-number-alignment=right]}
\toprule
& \multicolumn{4}{c}{{Parent Species}}\\
\cmidrule{2-5}
 \relax& {\ce{O}} \relax& {\ce{O2}} \relax& {\ce{H2O}} \relax& {\ce{CO2}}\\
\midrule
\SI{121.6}{nm} H\,\textsc{i} \relax& {---} \relax& {---} \relax& 4.53 \relax& {---} \\
\SI{130.4}{nm} O\,\textsc{i} \relax& 33.1 \relax& 1.29 \relax& 0.177 \relax& 0.0481 \\
\SI{135.6}{nm} O\,\textsc{i}] \relax& 0.539 \relax& 2.96 \relax& 0.0414 \relax& 0.0458 \\
\SI{297.2}{nm} [O\,\textsc{i}] \relax& 0.0598 \relax& 0.0598 \relax& 0.0598 \relax& 0.0598 \\
\SI{486.1}{nm} H\,\textsc{i} \relax& {---} \relax& {---} \relax& 0.402 \relax& {---} \\
\SI{557.7}{nm} [O\,\textsc{i}] \relax& 1 \relax& 1 \relax& 1 \relax& 1 \\
\SI{630.0}{nm} [O\,\textsc{i}] \relax& 2.40 \relax& 11.3 \relax& 0.943 \relax& 0.872 \\
\SI{636.4}{nm} [O\,\textsc{i}] \relax& 0.775 \relax& 3.64 \relax& 0.305 \relax& 0.282 \\
\SI{656.3}{nm} H\,\textsc{i} \relax& {---} \relax& {---} \relax& 2.24 \relax& {---} \\
\SI{777.4}{nm} O\,\textsc{i} \relax& 0.194 \relax& 2.06 \relax& 0.0609 \relax& 0.0288 \\
\SI{844.6}{nm} O\,\textsc{i} \relax& 4.43 \relax& 0.963 \relax& 0.181 \relax& 0.0209
\\
\bottomrule
\end{tabular}
\end{table}

\section{Conclusions}
In this study we presented a time-series analysis of 17 high-resolution spectra of Ganymede's auroral emission at optical wavelengths taken with Keck I/HIRES on \mbox{2021-06-08 UTC}. This study was the first to resolve and analyze the spatial variability of Ganymede's aurora at optical wavelengths and the first at any wavelength to evaluate these changes on a cadence of just a few minutes. We observed Ganymede during eclipse by Jupiter, which allowed us to capture the aurora without the overwhelming presence of reflected solar continuum. The timing of the eclipse also allowed us to observe Ganymede as it passed through the mid-plane of the Jovian plasma sheet where it was subjected to the highest incident electron number density. The high cadence of the observations also let us evaluate the potential for the rapid collapse of a localized \ce{H2O} atmosphere near the sub-solar point as Ganymede passed into Jupiter's shadow.

We observed Ganymede's plasma sheet mid-plane-facing hemisphere brightening relative to the opposite hemisphere, reaching a peak hemispheric brightness ratio of nearly 2 when it was about \qty{1}{\RJ} from the centrifugal equator of the plasma sheet. We did not observe the same exponential drop-off of brightness with distance from the plasma sheet mid-plane that has been seen at Europa \citep{Roth2016,deKleer2023}, though there was additional variability in the brightness which we attributed to longitudinal density heterogeneities in the Jovian plasma sheet. Additional observations could help to provide a better understanding of the nature of the variability seen in the brightness.

In addition to evaluating the north-south hemispheric brightness ratio, we also compared the dusk-dawn (trailing-leading) hemispheric brightness ratio. This ratio did not change with plasma sheet distance like the north-south ratio, but we did find the dusk hemisphere was almost always nearly two-times brighter than the dawn hemisphere. This effect is likely a combination of both higher incident electron number density on the ram-facing trailing hemisphere and a higher \ce{O2} column density toward dusk due to higher afternoon surface temperatures predicted by models \citep[e.g.,][]{Leblanc2017,Oza2018}. Unfortunately we cannot decouple these effects with eclipse observations because of the viewing geometry limitations.

Our MCMC model found an median atmospheric $\ce{H2O}/\ce{O2}$ column density ratio of $0.10${\raisebox{0.5ex}{\tiny$^{+0.05}_{-0.04}$}}, matching the upper limit of 0.10 found by \cite{deKleer2023} for \mbox{2021-06-08 UTC} but well below the \ce{H2O} abundance found by \cite{Roth2021} from sunlit UV observations. The $\qty{630.0}{nm}/\qty{557.7}{nm}$ brightness ratio is well suited to differentiating between electron impact on \ce{O2} and \ce{H2O} as the source of the excited \ce{O} atoms producing the auroral emission. We quantitatively tested our ability to detect an \ce{H2O} atmosphere localized to near the sub-solar point by using the Monte Carlo simulations of \cite{Leblanc2023} for Ganymede's \ce{O2} and \ce{H2O} atmospheres in eclipse. We combined these simulations with our aurora model to simulate emission components proposed by \cite{Roth2021} and evaluate their detectability with HIRES.

The modeled emission from the simulated sunlit \ce{H2O} and \ce{O2} atmospheres produced brightnesses of \qty{33}{R} at \qty{557.7}{nm} and \qty{45}{R} at \qty{656.3}{nm}, well above our typical observed brightnesses of \qtylist{11.5(0.6);0(2)}{R}, respectively. Our results are therefore inconsistent with the presence of a localized high-density \ce{H2O} atmosphere near the disk center proposed by \cite{Roth2021} based on sunlit observations. 

Even though the simulations of the eclipse atmospheres suggest the \ce{H2O} column density decreases rapidly during eclipse ingress, emission from electron impact on the actively-condensing \ce{H2O} atmosphere should still be a significant component of the observed \qtylist{557.7;656.3}{nm} brightnesses for the first two observations in our data set. We did not detect any variability in the brightness which would suggest an actively-condensing \ce{H2O} atmosphere during eclipse ingress, implying that if a sublimation \ce{H2O} atmosphere exists in sunlight and freezes back onto the surface during eclipse, the timescale for condensation must be around 10 minutes or less.

KdK acknowledges support from NASA through a grant to program HST-GO-15425 from the Space Telescope Science Institute, which is operated by the Associations of Universities for Research in Astronomy, Inc., under NASA contract NAS5-26555. FL acknowledges support by \textit{l'Agence Nationale de la Recherche} (ANR) under projects ANR-22-CE49-005-002 and ANR-21-CE49-0019. CS gratefully acknowledges the auspices of NASA's Solar System Observations program under contract 80NSSC22K0954. 

This work benefited from scientific exchanges that took place within International Space Sciences Institute (ISSI) international Team \#559 and as part of an ISSI workshop, Team \#515.

The data presented herein were obtained at the W.~M. Keck Observatory, which is operated as a scientific partnership among the California Institute of Technology, the University of California and the National Aeronautics and Space Administration. The Observatory was made possible by the generous financial support of the W.~M. Keck Foundation.

This research has made use of the Keck Observatory Archive (KOA), which is operated by the W.~M. Keck Observatory and the NASA Exoplanet Science Institute (NExScI), under contract with the National Aeronautics and Space Administration.

The authors wish to recognize and acknowledge the very significant cultural role and reverence that the summit of Maunakea has always had within the indigenous Hawaiian community.  We are most fortunate to have the opportunity to conduct observations from this mountain.

\facilities{W.~M. Keck Observatory (Keck I/HIRES)}

\bibliography{references}{}
\bibliographystyle{aasjournal}

\appendix

\section{Data Files}
Table \ref{tab:files} lists the file names and corresponding observation type for each FITS file used in this study. All data are available from the Keck Observatory Archive (KOA)\footnote{\url{http://koa.ipac.caltech.edu}}. These data were taken as a part of program ID C294 with principal investigator Katherine de~Kleer.

\LTcapwidth=\columnwidth
\begin{longtable}{lll}
\caption{Data files used in this study and their corresponding observation type and target. All Ganymede observations were taken during eclipse.}\label{tab:files}\\
\endfirsthead
\toprule
KOA Unique File Name \relax& Type \relax& Target\\
\midrule
\texttt{HI.20210608.09072.fits.gz} \relax& Calibration \relax& None (bias)\\
\texttt{HI.20210608.09116.fits.gz} \relax& Calibration \relax& None (bias)\\
\texttt{HI.20210608.09160.fits.gz} \relax& Calibration \relax& None (bias)\\
\texttt{HI.20210608.09205.fits.gz} \relax& Calibration \relax& None (bias)\\
\texttt{HI.20210608.09249.fits.gz} \relax& Calibration \relax& None (bias)\\
\texttt{HI.20210608.09293.fits.gz} \relax& Calibration \relax& None (bias)\\
\texttt{HI.20210608.09338.fits.gz} \relax& Calibration \relax& None (bias)\\
\texttt{HI.20210608.09382.fits.gz} \relax& Calibration \relax& None (bias)\\
\texttt{HI.20210608.09426.fits.gz} \relax& Calibration \relax& None (bias)\\
\texttt{HI.20210608.09471.fits.gz} \relax& Calibration \relax& None (bias)\\
\texttt{HI.20210608.09540.fits.gz} \relax& Calibration \relax& Quartz flat lamp\\
\texttt{HI.20210608.09586.fits.gz} \relax& Calibration \relax& Quartz flat lamp\\
\texttt{HI.20210608.09630.fits.gz} \relax& Calibration \relax& Quartz flat lamp\\
\texttt{HI.20210608.09676.fits.gz} \relax& Calibration \relax& Quartz flat lamp\\
\texttt{HI.20210608.09746.fits.gz} \relax& Calibration \relax& \ce{ThAr} arc lamp\\
\texttt{HI.20210608.09791.fits.gz} \relax& Calibration \relax& \ce{ThAr} arc lamp\\
\texttt{HI.20210608.09838.fits.gz} \relax& Calibration \relax& \ce{ThAr} arc lamp\\
\texttt{HI.20210608.09883.fits.gz} \relax& Calibration \relax& \ce{ThAr} arc lamp\\
\texttt{HI.20210608.09928.fits.gz} \relax& Calibration \relax& \ce{ThAr} arc lamp\\
\midrule
\texttt{HI.20210608.46696.fits.gz} \relax& Science \relax& Ganymede\\
\texttt{HI.20210608.47042.fits.gz} \relax& Science \relax& Io\\
\texttt{HI.20210608.47128.fits.gz} \relax& Science \relax& Ganymede\\
\texttt{HI.20210608.47493.fits.gz} \relax& Science \relax& Io\\
\texttt{HI.20210608.48132.fits.gz} \relax& Science \relax& Ganymede\\
\texttt{HI.20210608.48488.fits.gz} \relax& Science \relax& Europa\\
\texttt{HI.20210608.48580.fits.gz} \relax& Science \relax& Ganymede\\
\texttt{HI.20210608.48934.fits.gz} \relax& Science \relax& Europa\\
\texttt{HI.20210608.49005.fits.gz} \relax& Science \relax& Ganymede\\
\texttt{HI.20210608.49366.fits.gz} \relax& Science \relax& Europa\\
\texttt{HI.20210608.49473.fits.gz} \relax& Science \relax& Ganymede\\
\texttt{HI.20210608.49830.fits.gz} \relax& Science \relax& Europa\\
\texttt{HI.20210608.49901.fits.gz} \relax& Science \relax& Ganymede\\
\texttt{HI.20210608.50265.fits.gz} \relax& Science \relax& Europa\\
\texttt{HI.20210608.50332.fits.gz} \relax& Science \relax& Ganymede\\
\texttt{HI.20210608.50687.fits.gz} \relax& Science \relax& Europa\\
\texttt{HI.20210608.50888.fits.gz} \relax& Science \relax& Ganymede\\
\texttt{HI.20210608.51252.fits.gz} \relax& Science \relax& Europa\\
\texttt{HI.20210608.51328.fits.gz} \relax& Science \relax& Ganymede\\
\texttt{HI.20210608.51689.fits.gz} \relax& Science \relax& Europa\\
\texttt{HI.20210608.51765.fits.gz} \relax& Science \relax& Ganymede\\
\texttt{HI.20210608.52125.fits.gz} \relax& Science \relax& Europa\\
\texttt{HI.20210608.52195.fits.gz} \relax& Science \relax& Ganymede\\
\texttt{HI.20210608.52554.fits.gz} \relax& Science \relax& Europa\\
\texttt{HI.20210608.52655.fits.gz} \relax& Science \relax& Ganymede\\
\texttt{HI.20210608.53016.fits.gz} \relax& Science \relax& Europa\\
\texttt{HI.20210608.53087.fits.gz} \relax& Science \relax& Ganymede\\
\texttt{HI.20210608.53444.fits.gz} \relax& Science \relax& Europa\\
\texttt{HI.20210608.53553.fits.gz} \relax& Science \relax& Ganymede\\
\texttt{HI.20210608.53987.fits.gz} \relax& Science \relax& Europa\\
\texttt{HI.20210608.54077.fits.gz} \relax& Science \relax& Ganymede\\
\texttt{HI.20210608.54443.fits.gz} \relax& Science \relax& Europa\\
\texttt{HI.20210608.54508.fits.gz} \relax& Science \relax& Ganymede\\
\texttt{HI.20210608.55199.fits.gz} \relax& Science \relax& Europa\\
\texttt{HI.20210608.55271.fits.gz} \relax& Science \relax& Jupiter\\
\bottomrule
\end{longtable}

\end{document}